\documentclass[journal]{IEEEtran}
\usepackage{lipsum}
\usepackage{color}
\usepackage{cite}
\ifCLASSINFOpdf
\usepackage[pdftex]{graphicx}
\usepackage{epstopdf}
\usepackage{amsthm}
\usepackage{multirow}
\graphicspath{{Images/}}
\usepackage{amsmath,amssymb,amsfonts}
\usepackage{cuted}
\ifCLASSOPTIONcompsoc
\usepackage[caption=false,font=normalsize,labelfont=sf,textfont=sf]{subfig}
\else
\usepackage[caption=false,font=footnotesize]{subfig}
\fi
\usepackage{stfloats}

\begin{document}
\title{$\mu$-synthesis-based Generalized Robust Framework for Grid-following and Grid-forming Inverters}
\author{Soham~Chakraborty,~\IEEEmembership{Student Member,~IEEE,}
        Sourav~Patel,~\IEEEmembership{Member,~IEEE,}
        and~Murti V. Salapaka,~\IEEEmembership{Fellow,~IEEE}% <-this % stops a space
\thanks{S. Chakraborty, S. Patel, M. V. Salapaka are with the Department of Electrical and Computer Engineering, University of Minnesota, Minneapolis, 55455 MN, USA~ e-mail: chakr138@umn.edu, patel292@umn.edu, murtis@umn.edu}% <-this % stops a space
% \thanks{Manuscript received April 19, 2005; revised August 26, 2015.}
\thanks{The authors acknowledge Advanced Research Projects Agency-Energy (ARPA-E) for supporting this research through the project titled ``Rapidly Viable Sustained Grid'' via grant no. DE-AR0001016.}
}
\maketitle
%%%%%%%%%%%%%%%%%%%%%%%%%START%%%%%%%%%%%%%%%%%%%%%%%%%%%%%%%%%%
\begin{abstract}
Grid-following and grid-forming inverters are integral components of microgrids and for integration of renewable energy sources with the grid. For grid following inverters, which need to emulate controllable current sources, a significant challenge is to address the large uncertainty of the grid impedance. For grid forming inverters, which need to emulate a controllable voltage source, large uncertainty due to varying loads has to be addressed. In this article, a $\boldsymbol{\mu}$-synthesis-based robust control design methodology, where performance under quantified uncertainty is guaranteed, is developed under a unified approach for both grid-following and grid-forming inverters. The control objectives, while designing the proposed optimal controllers, are: $\mathbf{i})$ reference tracking, disturbance rejection, harmonic compensation capability with sufficient $\mathbf{LCL}$ resonance damping under large variations of grid impedance uncertainty for grid-following inverters; $\mathbf{ii})$ reference tracking, disturbance rejection, harmonic compensation capability with enhanced dynamic response under large variations of equivalent loading uncertainty for grid-forming inverters. A combined system-in-the-loop (SIL), controller hardware-in-the-loop (CHIL) and power hardware-in-the-loop (PHIL) based experimental validation on $\mathbf{10}$ kVA microgrid system with two physical inverter systems is conducted in order to evaluate the efficacy and viability of the proposed controllers.
\end{abstract}
%%%%%%%%%%%%%%%%%%%%%%%%%END%%%%%%%%%%%%%%%%%%%%%%%%%%%%%%%%%%
%%%%%%%%%%%%%%%%%%%%%%%%%START%%%%%%%%%%%%%%%%%%%%%%%%%%%%%%%%%%
%%%%%%%%%%%%%%%%%%%%%%%%%END%%%%%%%%%%%%%%%%%%%%%%%%%%%%%%%%%%
%%%%%%%%%%%%%%%%%%%%%%%%%START%%%%%%%%%%%%%%%%%%%%%%%%%%%%%%%%%%
\section{Introduction}\label{intro}
\IEEEPARstart{W}{ith} the proliferation of inverter-interfaced distributed energy resources, there is a renewed emphasis on local microgrids that provide operational flexibility and aid sustainability for the energy infrastructures. Both grid-forming (GFM) and grid-following (GFL) inverters have become essential components that have pivotal roles to play in such microgrids operating both in grid-tied and islanded mode. During islanded mode, GFM inverters maintain a stable voltage and frequency of the microgrid in the absence of grid. Whereas, GFL inverters are usually operated to supply power where voltage and frequency are maintained either by the grid or other GFM inverters \cite{DER_control_Madureira}. In the hierarchical structure of microgrid control system, \textit{inner} voltage-control loops, regulating voltage to specified values, are responsible for GFM inverters to emulate controllable voltage sources. Similarly, GFL inverters emulate controllable current sources by regulating currents via \textit{inner} current-control loop \cite{bidram2012}. 
\par Typically GFL inverters are connected to grid via $\mathrm{LCL}$ filters for high frequency attenuation caused by switching. Multiple important factors are considered in the design stage of GFL inverters including $1)$ resonances caused by low-damping of the $\mathrm{LCL}$ filter in GFL inverters, while connected to weak grid, that could lead to system instability. Here, proper damping of such resonance is crucial while designing \cite{lcldamp2}; $2)$ Uncertainty in variation of grid impedance parameters significantly influence the robustness of the output current controller. For example, increase in grid inductance requires reduction in the gain and bandwidth of the current controller to keep the system stable that leads to degradation of tracking performance and disturbance rejection capability. Here, there is a need for control design that delivers optimal performance while guaranteeing stability for the range of grid impedance encountered in practice \cite{robustcon2}; $3)$ Grid impedance variation causes variation in resonance frequency of the inverter system that impacts active damping methods. Here, robustness of the active damping to remain effective under uncertainty is required \cite{gridimp1}; $4)$ Most importantly, the controller should result in good tracking performance, disturbance rejection and harmonic compensation capability while remaining implementable in low-cost micro-controllers. Various types of control schemes and their advancements for GFL inverters are proposed in literature \cite{GFLSOA}. Major current-control schemes for GFL inverters, reported in existing literature, are compared in Fig.~\ref{SOA}(a). In summary, existing methods include classical (i.e. proportional/integral/resonant controller-based), hysteresis, sliding-mode, model predictive, repetitive, $\mathrm{LQR}$, $\mathcal{H}_\infty$-based control schemes as reported in \cite{GFLSOA,lcldamp2,piprmod2,universal,statefeed1,classical_new,pipr2,pipr5,pipr8,newfilter,robustcon1,robustcon2,gridimp1,hyst1,repet1,modelpre1,sliding1,twodegree1,hornic1,chakrabortyGFL,gridimp2,gridimp5}. Broadly, these schemes use, either passive damping with or without new filter topology or, active damping using either additional measurements for feedforward control action. These additional measurements increase the cost of availing multiple sensors. Moreover, most of the schemes do not provide guaranteed robustness both in active damping and in controller performance under varying and uncertain grid impedance.
\par Similarly, designing the voltage controller for GFM inverter is essentially a multi-objective task. The design considerations include $1)$ reference tracking, disturbance rejection and harmonic compensation in presence of various linear and non-linear loads; $2)$ The voltage controllers are required to provide compensation to dynamic variations of the output load current and improve the dynamic response \cite{yazdani}; $3)$ Unknown nature of the output loading of GFM inverter can significantly alter the system behavior. Here during heavily loaded condition of GFM inverter, transient performance is severely compromised \cite{multiloopGFM2}. Therefore, the voltage controller should be robust enough against unknown loading uncertainties to perform all the aforementioned tasks. Numerous voltage control strategies are proposed in the literature during past decades for GFM inverter system \cite{GFMSOA}. Major voltage-control schemes, reported in existing literature, are shown in Fig.~\ref{SOA}(b). In summary, there are nested-loop classical (i.e. proportional/integral/resonant controller-based), sliding-mode, model predictive, repetitive, $\mathrm{LQR}$, $\mathcal{H}_\infty$-based, $\mathcal{H}_2$-based control schemes as reported in \cite{GFMSOA,multiloopGFM2,classical1,classical2,classical3,multiloop1,multiloop2,multiloop3,hinf1GFM,hinf2GFM,zhong1,multihinf1,multihinf2,multihinf3,multihinf4,chakrabortyGFM,h2hinf1,h2hinf2}. Most of the advancements have focused on either multiple nested-loop structures with advanced control techniques or adding extra measurements as feedforward for enhancing dynamic performance. This results in either increased complication of the control structure that lead to difficulties in implementation or increase in the capital cost for availing multiple sensors. Also, most of the schemes face deteriorating dynamic response and lack robustness in controller performance under varying and uncertain equivalent loading. 
\begin{figure*}[t]
\centering
% \captionsetup[subfigure]{labelformat=empty}
\scriptsize
\subfloat[Major works on grid-following inverter control]{\begin{tabular}{l|l|l|}
\hline
\multicolumn{1}{c|}{$\mathbf{Ref.}$} 
& 
\multicolumn{1}{c|}{\begin{tabular}[c]{@{}c@{}}$\mathbf{LCL}$\\ $\mathbf{Damping}$\end{tabular}} 
& 
\multicolumn{1}{c|}{\begin{tabular}[c]{@{}c@{}}$\mathbf{Robustness~Against}$\\  $\mathbf{Grid~Impedance}$ \\ $\mathbf{Parameter~Variation}$\end{tabular}} \\ \hline
\begin{tabular}[c]{@{}l@{}} 
\cite{lcldamp2}\\
\cite{piprmod2,universal,statefeed1,classical_new,pipr2,pipr5,pipr8,newfilter,robustcon1}
\end{tabular} 
&
\begin{tabular}[c]{@{}l@{}}
$\mathrm{\bullet~New~filter~topology}$\\
$\mathrm{\bullet~Additional~feedforward~of}$\\$\mathrm{grid~voltage,~grid~current,}$\\
$\mathrm{inner~capacitor~current}$\\
$\mathrm{\bullet~State~feedback~control}$\\
$\mathrm{\bullet~Lack~robust~damping}$
\end{tabular} 
& 
\begin{tabular}[c]{@{}l@{}}
$\mathrm{\bullet~Lack~robustness}$\\ $\mathrm{under~varying~unknown}$\\$\mathrm{grid~impedance}$\\
$\mathrm{\bullet~Designed~with}$\\$\mathrm{known,~fixed,~nominal}$\\$\mathrm{grid~impedance}$
\end{tabular} \\ \hline
\begin{tabular}[c]{@{}l@{}}
\cite{robustcon2,gridimp1}\\
\cite{hyst1,repet1,modelpre1,sliding1,twodegree1,hornic1}\\ 
\cite{chakrabortyGFL,gridimp2,gridimp5}
\end{tabular} 
& 
\begin{tabular}[c]{@{}l@{}}
$\mathrm{\bullet~Passive~damping}$\\
$\mathrm{\bullet~Additional~outer~current}$\\$\mathrm{inner~current~loop}$\\
$\mathrm{\bullet~Inherent~active~damping}$\\$\mathrm{using~controller~shaping}$\\
$\mathrm{\bullet~Challenging~during}$\\$\mathrm{varying~grid~impedance}$
\end{tabular} 
& 
\begin{tabular}[c]{@{}l@{}}
$\mathrm{\bullet~Lack~robustness}$\\ $\mathrm{under~varying~unknown}$\\$\mathrm{grid~impedance}$\\
$\mathrm{\bullet~Grid~impedance}$\\$\mathrm{uncertainty~not~designed}$\\$\mathrm{for~controller~synthesis}$
\end{tabular} \\ \hline
\begin{tabular}[c]{@{}l@{}}
$\mathrm{Propos}$-\\$\mathrm{ed}$
\end{tabular} 
& 
\begin{tabular}[c]{@{}l@{}}
$\mathrm{\checkmark~Single~loop~control~leads}$\\
$\mathrm{reduction~in~measurements}$\\
$\mathrm{\checkmark~Inherent~active~damping}$\\
$\mathrm{using~controller~shaping}$\\
$\mathrm{\checkmark~Robust~damping~under}$\\
$\mathrm{varying~grid~impedance}$
\end{tabular} 
& 
\begin{tabular}[c]{@{}l@{}}
$\mathrm{\checkmark~Grid~impedance}$\\
$\mathrm{uncertainty~designed}$\\
$\mathrm{for~controller~synthesis}$\\ 
$\mathrm{\checkmark~Exhibit~robustness}$\\ $\mathrm{under~varying~unknown}$\\$\mathrm{grid~impedance}$
\end{tabular} \\ \hline
\end{tabular}}~ %%%%%%%%%%%%%%%%%%%%%%%%%%%%%% GFM
\subfloat[Major works on grid-forming inverter control]{\begin{tabular}{|l|l|l}
\hline
\multicolumn{1}{|c|}{$\mathbf{Ref.}$} 
& 
\multicolumn{1}{c|}{\begin{tabular}[c]{@{}c@{}}$\mathbf{Improved}$\\ $\mathbf{Dynamic~Response}$\end{tabular}} 
& 
\multicolumn{1}{c}{\begin{tabular}[c]{@{}c@{}}$\mathbf{Robustness~Against}$\\  $\mathbf{Equivalent~Load}$ \\ $\mathbf{Parameter~Variation}$\end{tabular}} \\ \hline
\begin{tabular}[c]{@{}l@{}} 
\cite{multiloopGFM2}\\
\cite{classical1,classical2,classical3}\\
\cite{multiloop1,multiloop2,multiloop3}
\end{tabular} 
&
\begin{tabular}[c]{@{}l@{}}
$\mathrm{\bullet~Multiloop~control~structure}$\\
$\mathrm{\bullet~Feedback~of~inductor~current,}$\\
$\mathrm{inner~capacitor~current}$\\
$\mathrm{\bullet~Disturbance~observer~design}$\\
$\mathrm{\bullet~Dominant~pole~elimination}$\\
$\bullet~\mathrm{Need~multiple~measurements}$
\end{tabular} 
& 
\begin{tabular}[c]{@{}l@{}}
$\mathrm{\bullet~Lack~robustness}$\\ $\mathrm{during~variation~of}$\\$\mathrm{equivalent~loading}$\\
$\mathrm{\bullet~Designed~controller~for}$\\$\mathrm{rated~loading~condition}$\\
$\mathrm{\bullet~No~loading~uncertainty}$
\end{tabular} \\ \hline
\begin{tabular}[c]{@{}l@{}}
\cite{hinf1GFM,hinf2GFM,zhong1}\\
\cite{multihinf1,multihinf2,multihinf3,multihinf4}\\ 
\cite{chakrabortyGFM,h2hinf1,h2hinf2}
\end{tabular} 
& 
\begin{tabular}[c]{@{}l@{}}
$\mathrm{\bullet~Multiloop~control~structure}$\\
$\mathrm{with~sliding~mode,~LQR,}$\\
$\mathrm{\mathcal{H}_2/\mathcal{H}_\infty,\mathcal{H}_\infty/\mathcal{H}_\infty~control}$\\
$\mathrm{\bullet~Complicated,~loopy~structure}$\\
$\mathrm{with~multiple~measurements}$\\
$\mathrm{\bullet~Challenging~during~varying}$\\$\mathrm{equivalent~loading~conditions}$
\end{tabular} 
& 
\begin{tabular}[c]{@{}l@{}}
$\mathrm{\bullet~Lack~robustness~for}$\\ $\mathrm{full~range~variation~of}$\\$\mathrm{equivalent~loading}$\\
$\mathrm{\bullet~Lack~robust~performance}$\\
$\mathrm{\bullet~Designed~controller~for}$\\$\mathrm{simple~load~uncertainty,}$\\
$\mathrm{without~non}$-$\mathrm{linear~load}$
\end{tabular} \\ \hline
\begin{tabular}[c]{@{}l@{}}
$\mathrm{Propos}$-\\$\mathrm{ed}$
\end{tabular} 
& 
\begin{tabular}[c]{@{}l@{}}
$\mathrm{\checkmark~\mathcal{H}_\infty}$-$\mathrm{based~robust~control}$\\
$\mathrm{\checkmark~Single~loop~control~structure}$\\
$\mathrm{\checkmark~No~need~of~feedforward~loop}$\\
$\mathrm{\checkmark~Reduction~in~measurements}$\\
$\mathrm{\checkmark~Enhancement~of~dynamic}$\\
$\mathrm{response~by~controller~shaping}$
\end{tabular} 
& 
\begin{tabular}[c]{@{}l@{}}
$\mathrm{\checkmark~Equivalent~loading}$\\
$\mathrm{uncertainty~designed}$\\
$\mathrm{for~controller~synthesis}$\\ 
$\mathrm{\checkmark~Exhibit~robustness}$\\ $\mathrm{under~the~variation~of}$\\
$\mathrm{equivalent~loading}$
\end{tabular} \\ \hline
\end{tabular}}
\caption{A summary of major current-control and voltage-control strategies for grid-following and grid-forming inverters proposed since last decade respectively.}
\label{SOA}
\end{figure*}
\par This article presents a generalized $\mu$-synthesis-based robust control framework and leverages a voltage-current duality in the plant dynamic model of GFL and GFM inverters. By modeling and quantifying the uncertainties in grid impedance parameters and uncertainties in equivalent loading parameters for GFL and GFM inverters respectively, the generalized control framework results in controllers that are single-loop, hence simple and cost-effective. The resulting controllers offer ease of implementation and provide optimal robustness in performance under uncertainties with respect to good reference tracking, disturbance rejection and harmonic compensation capabilities. Moreover, the resulting current-controller for GFL inverter provides inherent active damping under grid parameter variation and the resulting voltage-controller for GFM inverter enhances the dynamic performance during load transients. A combined system-in-the-loop (SIL), controller-hardware-in-the-loop (CHIL) and power-hardware-in-the-loop (PHIL) -based experiment is conducted to evaluate the efficacy and viability of the proposed controllers.
\par This article is organized as follows. In Section \ref{motivation}, the problem formulation and motivation of the study are presented. In Section \ref{control}, the proposed generalized $\mu$-synthesis-based robust control framework is described. Section \ref{controller} provides the controller synthesis and corresponding stability analysis. In Section \ref{result}, the experimental setup and corresponding results are described. Finally, Section \ref{conclusion} concludes the paper.
%%%%%%%%%%%%%%%%%%%%%%%%%END%%%%%%%%%%%%%%%%%%%%%%%%%%%%%%%%%%
%%%%%%%%%%%%%%%%%%%%%%%%%START%%%%%%%%%%%%%%%%%%%%%%%%%%%%%%%%%%
\begin{figure}[t]
	\centering
	\subfloat[]{\includegraphics[scale=0.163,trim={0cm 0cm 25cm 5cm},clip]{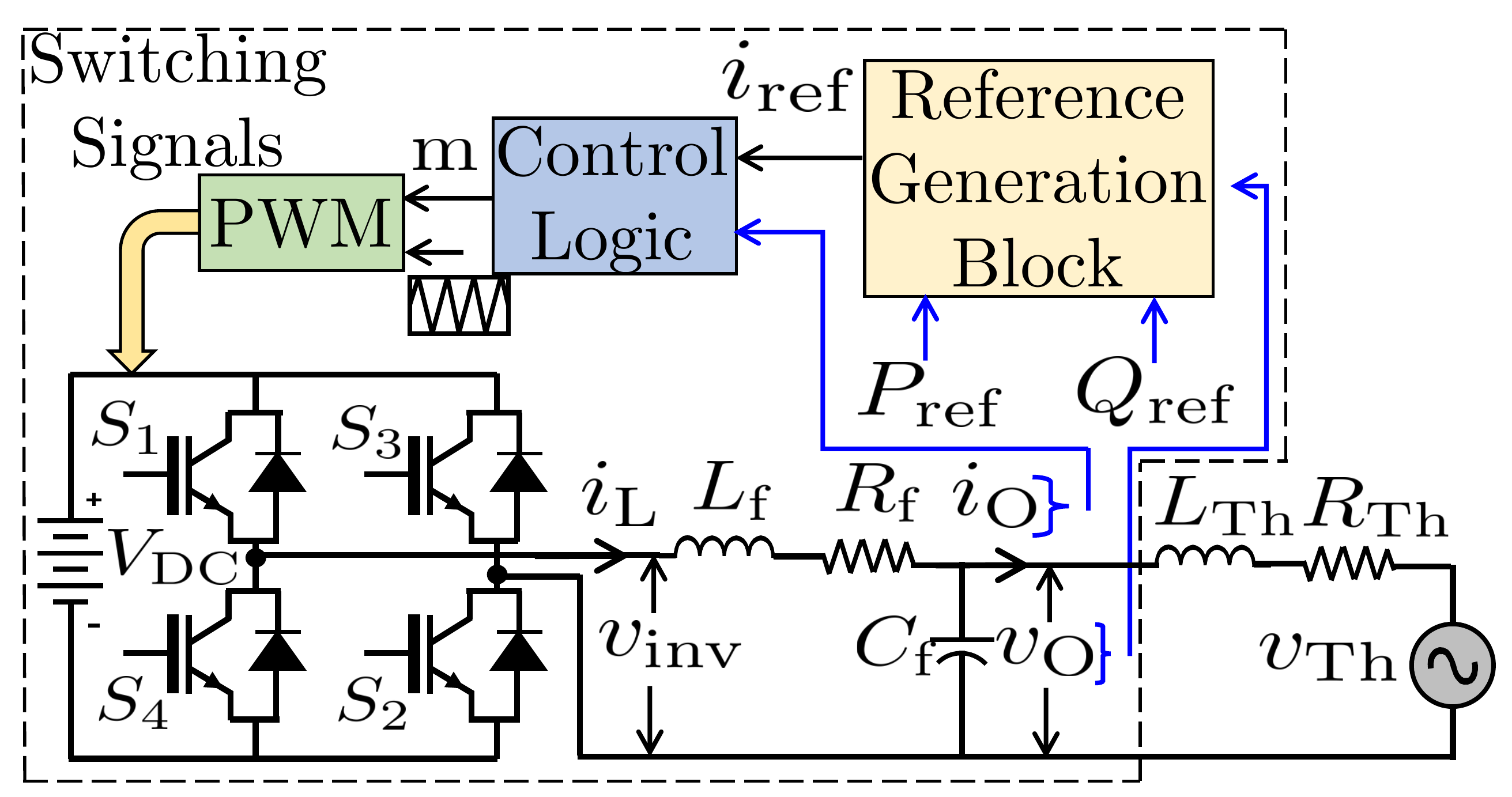}%
	\label{fig:GFLcktv1}}~
	\subfloat[]{\includegraphics[scale=0.163,trim={0cm 0cm 25cm 5cm},clip]{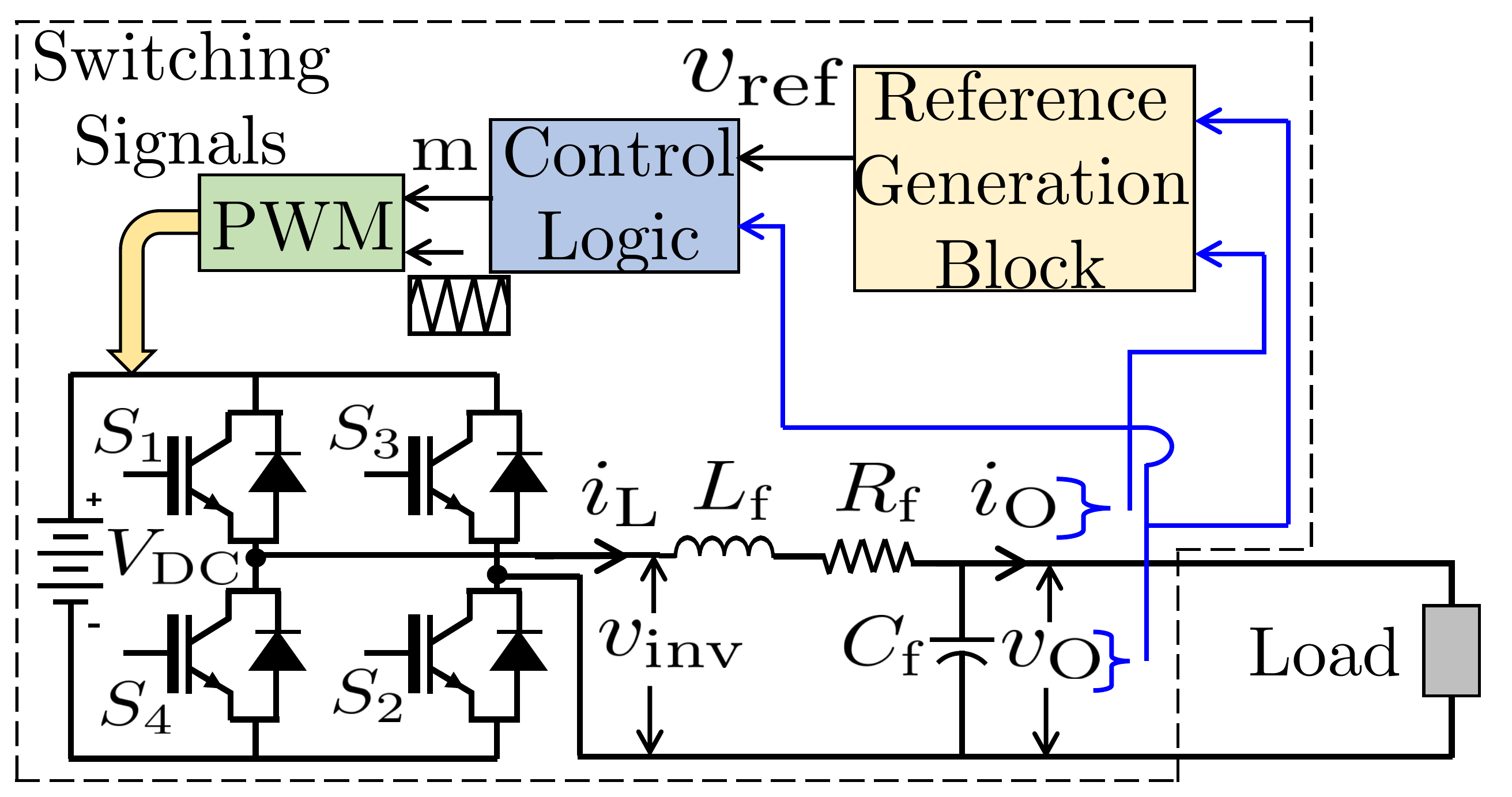}%
	\label{fig:GFMcktv1}}
	\caption{A $1$-$\phi$ (a) grid-following inverter, and (b) grid-forming inverter.}\label{fig:GFLGFMckt}
\end{figure}
\section{Problem Formulation and Motivation}\label{motivation}
In this section control problem formulation will be discussed for inverter system operating in GFL and GFM mode. For both GFL and GFM inverters, a single-phase H-bridge inverter is considered comprising a dc bus, $V_\mathrm{DC}$, four switching devices, $S_1, S_2, S_3, S_4$, and an $\mathrm{LCL}$ filter with $L_\mathrm{f}$, $L_\mathrm{g}$, $R_\mathrm{f}$, $R_\mathrm{g}$ and $C_\mathrm{f}$ as inverter and grid side filter inductors, parasitic resistances and filter capacitor respectively. A control layer is employed with sinusoidal PWM switching technique to generate the switching signals for power circuit.
%%%%%%%%%%%%%%%%%%%%%%%%%%%%%%%%%%%%%%%%%%%%%%%%%%%%%%%%%%%%%%%%%%%%
\subsection{Problem Formulation for Grid-following Inverter}\label{probGFL}
% \begin{figure}[t]
% 	\centering
% 	\includegraphics[scale=0.23,trim={0cm 0cm 25cm 5cm},clip]{GFL_ckt_v2.pdf}%
% 	\caption{\color{red}adjust figure\color{black}A $1$-$\phi$ grid-following inverter connected to distribution network.}
% 	\label{fig:GFLcktv1}
% \end{figure}
A GFL inverter is connected to a distribution grid, represented by the Thevenin equivalent voltage source, $v_\mathrm{Th}$, in series with the Thevenin equivalent impedance, $Z_\mathrm{Th}:=R_\mathrm{Th}+j\omega_\mathrm{o}L_\mathrm{Th}$. $\omega_\mathrm{o}$ (in rad/sec) is the frequency of the distribution network. For simplicity, the Thevenin equivalent impedance accounts grid side filter parameters also (i.e. $L_\mathrm{g}$, $R_\mathrm{g}$). Various components of GFL inverter system are shown in Fig.~\ref{fig:GFLGFMckt}\subref{fig:GFLcktv1}. In this operation, the inverter operates with an output current control strategy to regulate real and reactive power output where the voltage and frequency of the distribution network are determined by another source such as the grid or other GFM inverters. The goal of the current controller is to generate a controlled voltage signal, $v_\mathrm{inv}$, by switching signals such that the output current, $i_\mathrm{O}$, tracks the reference signal, $i_\mathrm{ref}$, generated by reference generation block as discussed in Appendix~$\mathrm{A}$. The dynamics of the inverter are described as:
\begin{align}
L_\mathrm{f} \frac{\mathrm{d}\langle i_\mathrm{L}\rangle}{\mathrm{d}t} +R_\mathrm{f}\langle i_\mathrm{L}\rangle &= \langle v_\mathrm{inv}\rangle - \langle v_\mathrm{O}\rangle, \label{eq1}\\
L_\mathrm{Th} \frac{\mathrm{d}\langle i_\mathrm{O}\rangle}{\mathrm{d}t} +R_\mathrm{Th}\langle i_\mathrm{O}\rangle &= \langle v_\mathrm{O}\rangle - \langle v_\mathrm{Th}\rangle, \label{eq2}\\
C_\mathrm{f} \frac{\mathrm{d}\langle v_\mathrm{O}\rangle}{\mathrm{d}t} &=  \langle i_\mathrm{L}\rangle - \langle i_\mathrm{O}\rangle, \label{eq3}
\end{align}
\noindent where $\langle .\rangle$ signifies the average values of the corresponding variable over one switching cycle ($T_\mathrm{s}$) \cite{yazdani}. Laplace transformation and algebraic manipulation with \eqref{eq1}, \eqref{eq2} and \eqref{eq3} result:
\begin{align}\label{eq4}
i_\mathrm{O}(s) &= \mathcal{G}_\mathrm{inv}^\mathrm{GFL}(s)v_\mathrm{inv}(s) - \mathcal{G}_\mathrm{Th}^\mathrm{GFL}(s)v_\mathrm{Th}(s),
\end{align}
\begin{figure*}[t]
	\centering
	\subfloat[]{\includegraphics[scale=0.113,trim={0cm 0cm 12cm 0cm},clip]{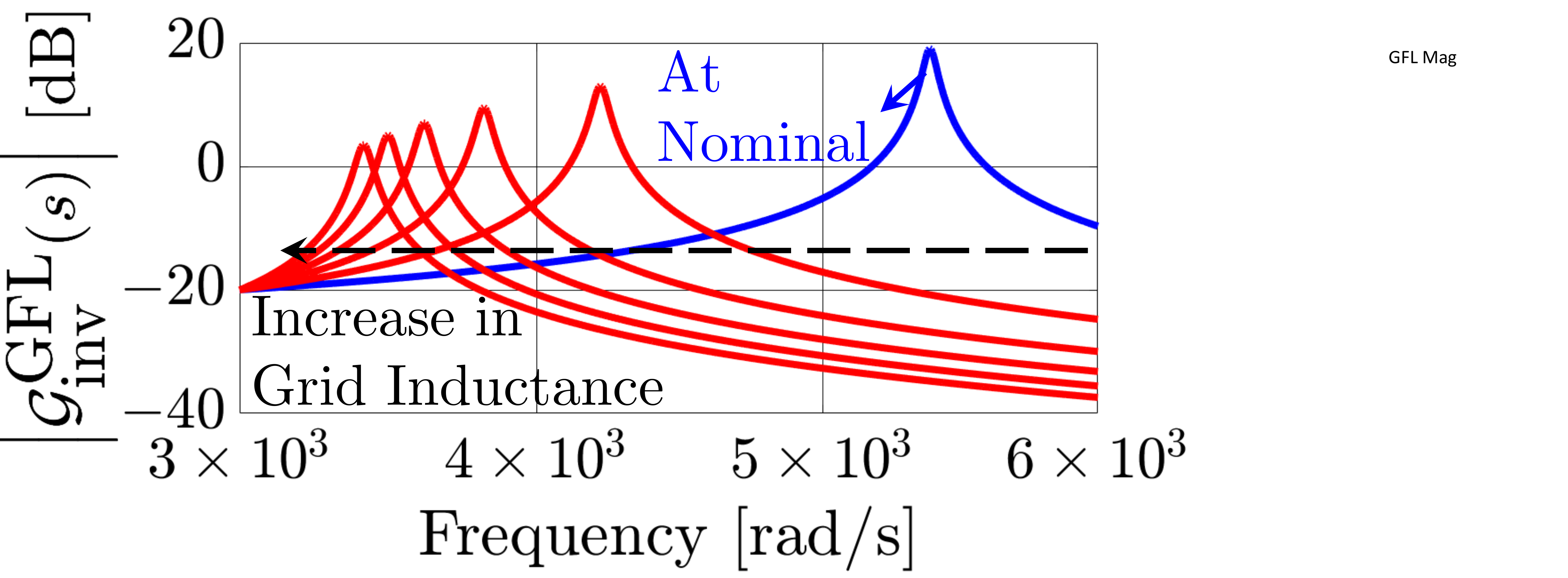}%
	\label{fig:GFL_mag}}~
	\subfloat[]{\includegraphics[scale=0.113,trim={0cm 0cm 12cm 0cm},clip]{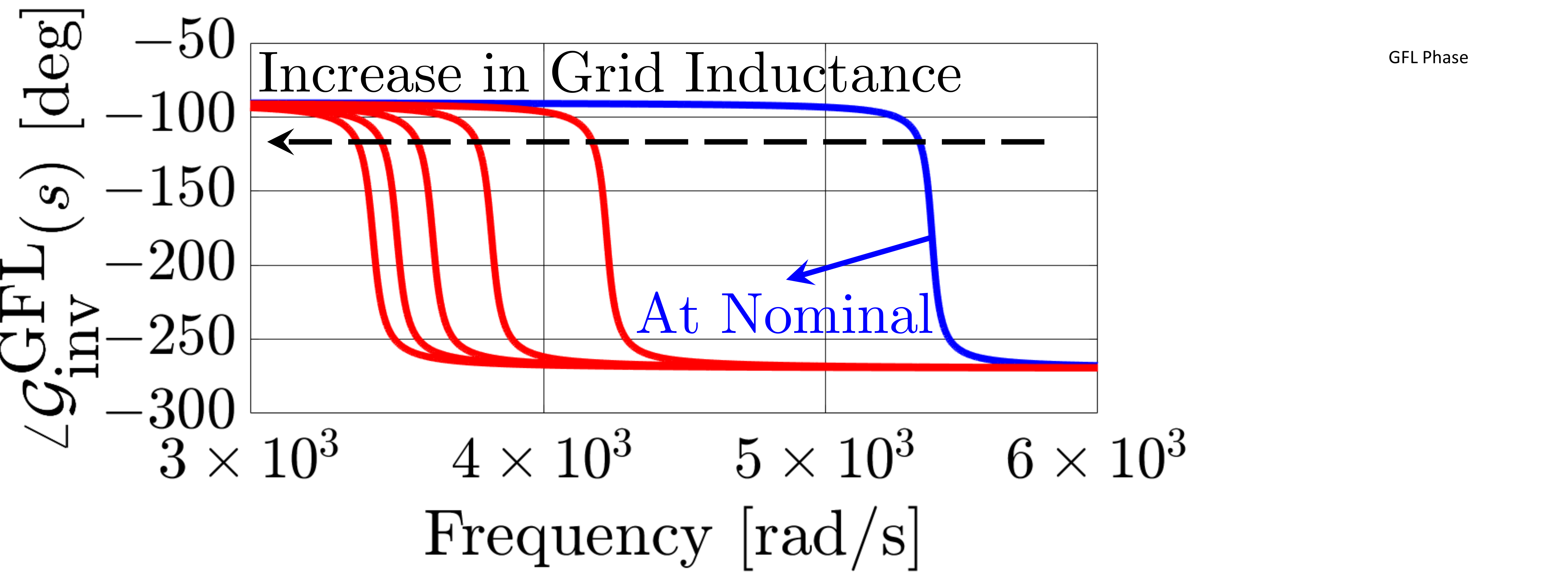}%
	\label{fig:GFL_phase}}~
	\subfloat[]{\includegraphics[scale=0.111,trim={0cm 0cm 13cm 0cm},clip]{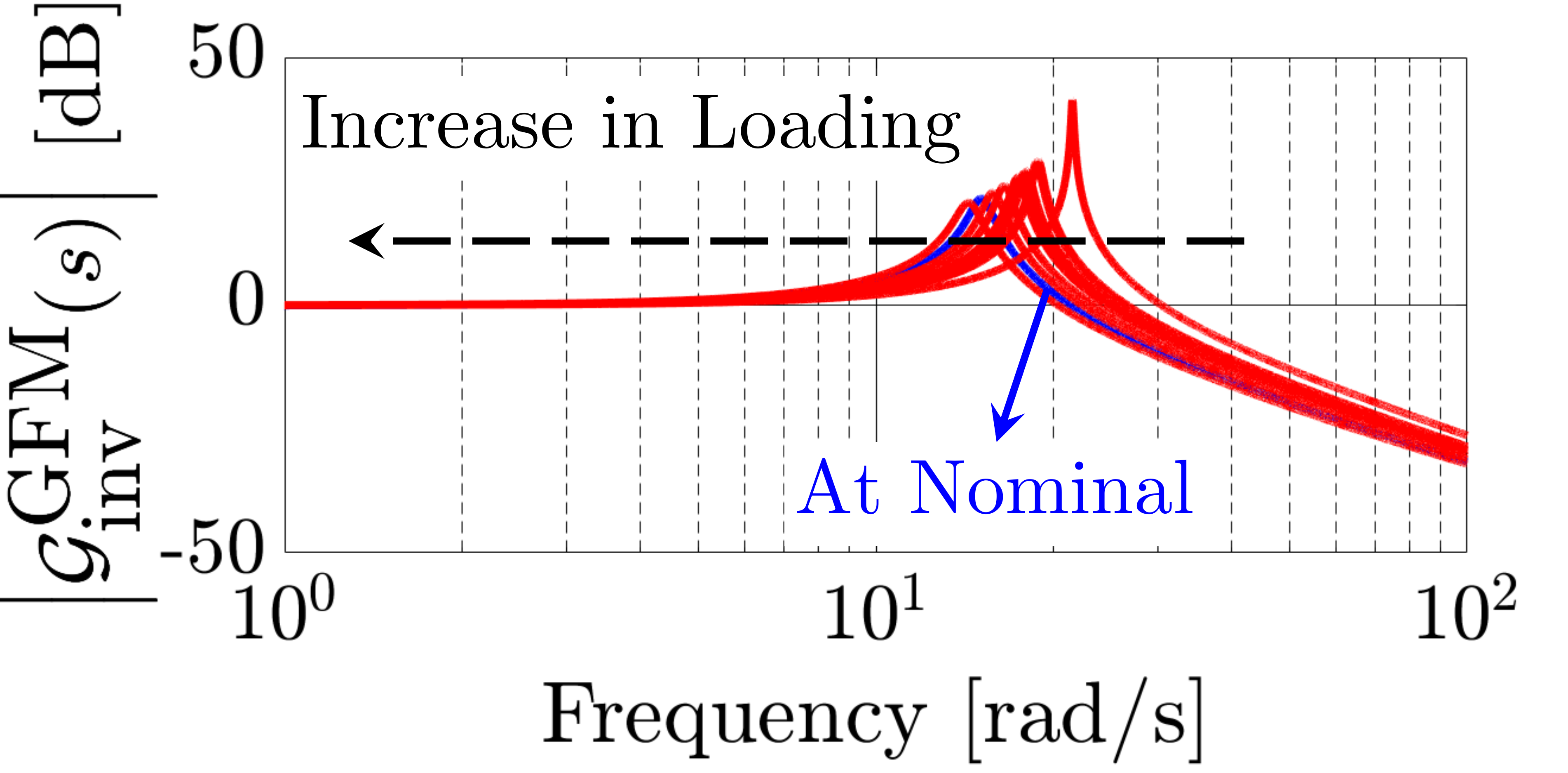}%
	\label{fig:GFM_mag}}~
	\subfloat[]{\includegraphics[scale=0.111,trim={0cm 0cm 13cm 0cm},clip]{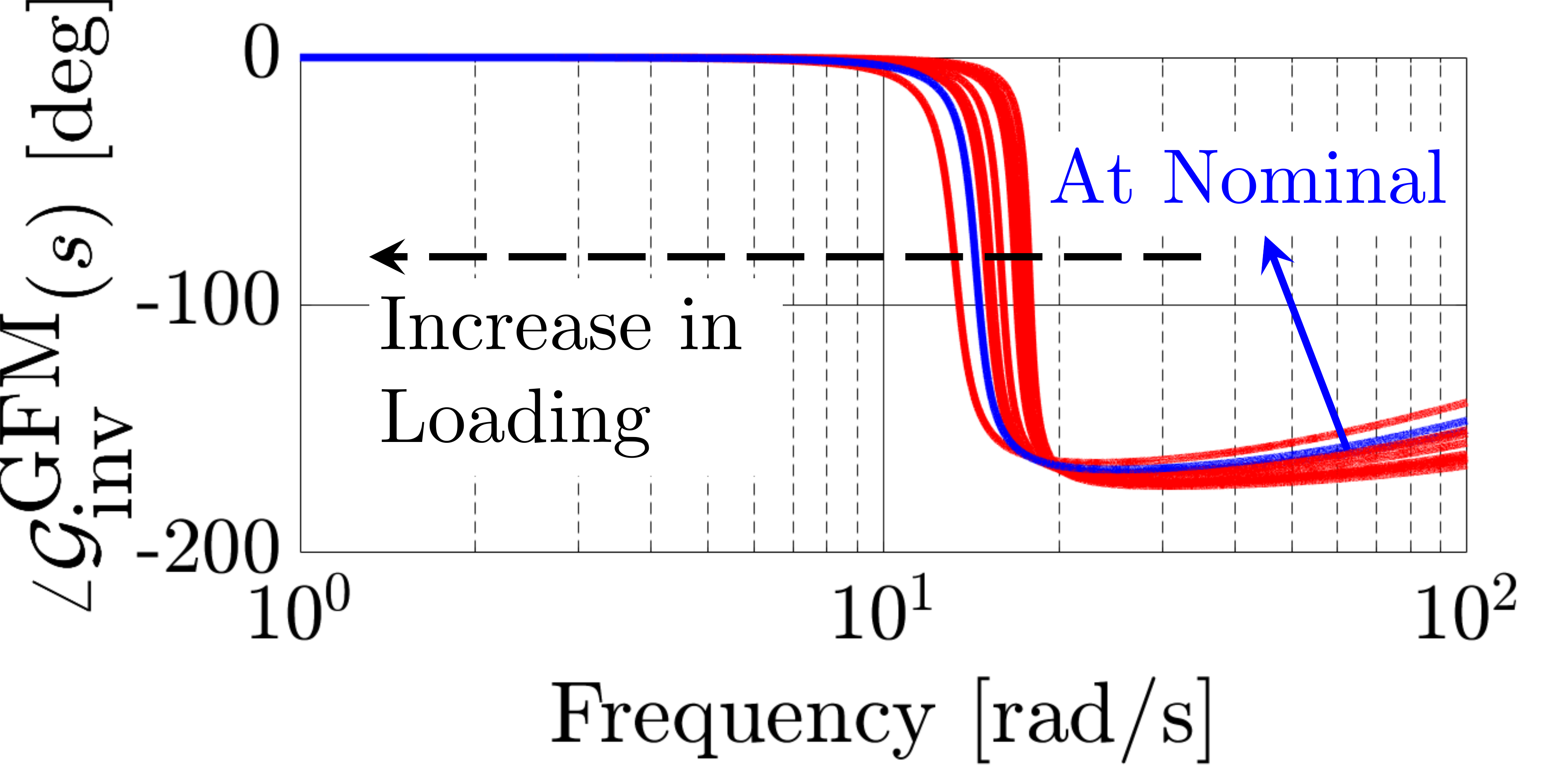}%
	\label{fig:GFM_phase}}
	\caption{Bode plots of (a) magnitude and (b) phase of $\mathcal{G}_\mathrm{inv}^\mathrm{GFL}(s)$  of \eqref{eq4} and (c) magnitude and (d) phase of $\mathcal{G}_\mathrm{inv}^\mathrm{GFM}(s)$ of \eqref{eq13} respectively.}
	\label{fig:GFLGFMmagphase}
\end{figure*}
%%%%%%%%%%%%%%%%%%%%%%%%%%%%%%%% 
\begin{figure*}[b]
% \scriptsize
\small
\begin{align}\hline \nonumber \\
\mathcal{G}_\mathrm{inv}^\mathrm{GFL}(s) &= {1}/{[(C_\mathrm{f}L_\mathrm{f}L_\mathrm{Th})s^3 + C_\mathrm{f}(L_\mathrm{Th}R_\mathrm{f}+L_\mathrm{f}R_\mathrm{Th})s^2 + (C_\mathrm{f}R_\mathrm{f}R_\mathrm{Th} + L_\mathrm{f} + L_\mathrm{Th})s + (R_\mathrm{f}+R_\mathrm{Th})}] \label{eq5}\\
\mathcal{G}_\mathrm{Th}^\mathrm{GFL}(s) &= {[(C_\mathrm{f}L_\mathrm{f})s^2 + (C_\mathrm{f}R_\mathrm{f})s + 1}]\mathcal{G}_\mathrm{inv}^\mathrm{GFL}(s)\label{eq6}
\end{align}
\end{figure*}
%%%%%%%%%%%%%%%%%%%%%%%%%%%%%%%%
\noindent where, $\mathcal{G}_\mathrm{inv}^\mathrm{GFL}(s)$ and $\mathcal{G}_\mathrm{Th}^\mathrm{GFL}(s)$ are transfer functions parameterized by $L_\mathrm{f}$, $R_\mathrm{f}$, $C_\mathrm{f}$, $L_\mathrm{Th}$ and $R_\mathrm{Th}$, as given in \eqref{eq5} and \eqref{eq6}. As observed in \eqref{eq4}, distribution network has two-fold impacts on the open-loop plant dynamics. $v_\mathrm{Th}(s)$ acts as an exogenous disturbance signal to the plant which can be addressed by classical classical disturbance rejection problem. Both $\mathcal{G}_\mathrm{inv}^\mathrm{GFL}(s)$ and $\mathcal{G}_\mathrm{Th}^\mathrm{GFL}(s)$ consist of $L_\mathrm{Th}$ and $R_\mathrm{Th}$ as parameters. Variation of these parameters due to changing distribution network topology introduces uncertainties in the plant model \cite{gridimp2,gridimp5}. Fig.~\ref{fig:GFLGFMmagphase}\subref{fig:GFL_mag} and Fig.~\ref{fig:GFLGFMmagphase}\subref{fig:GFL_phase} show the Bode plot of $\mathcal{G}_\mathrm{inv}^\mathrm{GFL}(s)$ that clearly shows the large variation in frequency response of the open-loop plant model due to grid inductance variation. Fig.~\ref{fig:GFLGFMmagphase}\subref{fig:GFL_mag} shows that the equivalent $\mathrm{LCL}$ resonant peak is sensitive to grid inductance, $L_\mathrm{Th}$. For stiff grid with small $L_\mathrm{Th}$ (leading to a large resonant frequency), the resultant resonant frequency is larger than the bandwidth of the controller. However, such a controller when used for a sufficiently weak grid with large $L_\mathrm{Th}$, the resultant resonance peak may enter the pass-band of the current controller that in turn results in instability \cite{gridimp2,gridimp5}. As a result, this uncertainty in grid impedance parameters introduces challenges with respect to performance under uncertainty of grid impedance. With this motivation, this article designs a single-loop $\mu$-synthesis-based stabilizing controller for GFL inverter that has robust active damping, tracking performance, disturbance rejection and harmonic compensation under grid impedance uncertainties.
% \begin{enumerate}
%     \item A single loop robust active damping for stabilizing the varying resonance peak.
%     \item Robust stable operation of the GFL VSI under wide range of grid impedance variation.
%     \item Robust tracking performance, disturbance rejection and selective harmonic compensation features of the controller with possibly higher grid impedance uncertainties.
% \end{enumerate}
%%%%%%%%%%%%%%%%%%%%%%%%%%%%%%%%%%%%%%%%%%%%%%%%%%%%%%%%%%%%%%%%%%%%
\subsection{Problem Formulation for Grid-forming Inverter}\label{probGFM}
% \begin{figure}[t]
% 	\centering
% 	\includegraphics[scale=0.23,trim={0cm 0cm 25cm 5cm},clip]{GFM_ckt_v2.pdf}%
% 	\caption{A $1$-$\phi$ grid-forming inverter connected across an electrical load.}
% 	\label{fig:GFMcktv1}
% \end{figure} 
Various components of a GFM inverter are shown in Fig.~\ref{fig:GFLGFMckt}\subref{fig:GFMcktv1}. A GFM inverter is connected across an equivalent load that represents the corresponding loading of the inverter while connected to any network. For simplicity, the grid side filter parameters are also included in the equivalent loading. In this mode, the inverter operates with an output voltage control strategy to generate a stable voltage and frequency across the equivalent load. The goal of the voltage control logic is to generate a controlled voltage signal, $v_\mathrm{inv}$, by switching signals such that the output voltage, $v_\mathrm{O}$, tracks the reference signal, $v_\mathrm{ref}$, generated by reference generation layer as discussed in Appendix~$\mathrm{B}$. The dynamics of the inverter are described as:
\begin{align}
L_\mathrm{f} \frac{\mathrm{d}\langle i_\mathrm{L}\rangle}{\mathrm{d}t} +R_\mathrm{f}\langle i_\mathrm{L}\rangle &= \langle v_\mathrm{inv}\rangle - \langle v_\mathrm{O}\rangle, \label{eq7}\\
C_\mathrm{f} \frac{\mathrm{d}\langle v_\mathrm{O}\rangle}{\mathrm{d}t} &=  \langle i_\mathrm{L}\rangle - \langle i_\mathrm{O}\rangle, \label{eq8}
\end{align}
where $\langle .\rangle$ signifies the average values of the corresponding variable over one switching cycle ($T_\mathrm{s}$) \cite{yazdani}. Laplace transformation and algebraic manipulation with \eqref{eq7} and \eqref{eq8} result:
\begin{align}\label{eq9}
v_\mathrm{O}(s) &= \mathcal{G}_\mathrm{1}(s)v_\mathrm{inv}(s) - \mathcal{G}_\mathrm{2}(s)i_\mathrm{O}(s),
\end{align}
where, $\mathcal{G}_\mathrm{1}(s)$ and $\mathcal{G}_\mathrm{2}(s)$ are transfer functions parameterized by $L_\mathrm{f}$, $R_\mathrm{f}$, $C_\mathrm{f}$, as given in \eqref{eq10} and \eqref{eq11} and described by: 
\begin{align}
\mathcal{G}_\mathrm{1}(s) &= \dfrac{1}{[L_\mathrm{f}C_\mathrm{f}]s^2 + [R_\mathrm{f}C_\mathrm{f}]s + 1}~\text{and} \label{eq10}\\ 
\mathcal{G}_\mathrm{2}(s) &= \dfrac{L_\mathrm{f}s+R_\mathrm{f}}{[L_\mathrm{f}C_\mathrm{f}]s^2 + [R_\mathrm{f}C_\mathrm{f}]s + 1}. \label{eq11}
\end{align}
The nature of the equivalent load is essential for characterizing $i_\mathrm{O}$. In this work, the equivalent load is modeled by a parallel combination of two components. First component is a linear admittance, $Y_\mathrm{Load}(s):=1/(L_\mathrm{Load}s+R_\mathrm{Load})$, with unknown $R_\mathrm{Load}$ and $L_\mathrm{Load}$ elements in series combination; the grid side filter parameters are included in $R_\mathrm{Load}$ and $L_\mathrm{Load}$. Another component is a parallel combination of current sources consisting of both fundamental and harmonic components \cite{loadmodel1} and defined as: $i_\mathrm{h}(s):=\sum_{\mathrm{k}}i_\mathrm{k}(s)$, where $\mathrm{k}$ is odd and $i_\mathrm{k}(s)$ is $\mathrm{k}^\mathrm{th}$ harmonic current. As a result, $i_\mathrm{O}(s)$ is characterized as:
\begin{align}\label{eq12}
    i_\mathrm{O}(s) &= Y_\mathrm{Load}(s)v_\mathrm{O}(s) + i_\mathrm{h}(s).
\end{align}
Combining \eqref{eq9} and \eqref{eq12}, system's dynamic equation becomes:
\begin{align}\label{eq13}
   v_\mathrm{O}(s) &= \mathcal{G}_\mathrm{inv}^\mathrm{GFM}(s)v_\mathrm{inv}(s) - \mathcal{G}_\mathrm{Load}^\mathrm{GFM}(s)i_\mathrm{h}(s),
\end{align}
where, $\mathcal{G}_\mathrm{inv}^\mathrm{GFM}(s)$ and $\mathcal{G}_\mathrm{Load}^\mathrm{GFM}(s)$ are transfer functions parameterized by $L_\mathrm{f}$, $R_\mathrm{f}$, $C_\mathrm{f}$, $L_\mathrm{Load}$ and $R_\mathrm{Load}$, as given in \eqref{eq14} and \eqref{eq15}.
%%%%%%%%%%%%%%%%%%%%%%%%%%%%%%%%
\begin{figure*}[b]
% \scriptsize
\small
\begin{align}\hline \nonumber\\
\mathcal{G}_\mathrm{inv}^\mathrm{GFM}(s) &= {[L_\mathrm{Load}s + R_\mathrm{Load}]}/{[(C_\mathrm{f}L_\mathrm{f}L_\mathrm{Load})s^3 + C_\mathrm{f}(L_\mathrm{Load}R_\mathrm{f}+L_\mathrm{f}R_\mathrm{Load})s^2 + (C_\mathrm{f}R_\mathrm{f}R_\mathrm{Load} + L_\mathrm{f} + L_\mathrm{Load})s + (R_\mathrm{f}+R_\mathrm{Load})]} \label{eq14}\\
\mathcal{G}_\mathrm{Load}^\mathrm{GFM}(s) &= \mathcal{G}_\mathrm{inv}^\mathrm{GFM}(s){[(L_\mathrm{Load}L_\mathrm{f})s^2 + (L_\mathrm{Load}R_\mathrm{f}+L_\mathrm{f}R_\mathrm{Load})s + (R_\mathrm{Load}R_\mathrm{f})]}/[L_\mathrm{Load}s + R_\mathrm{Load}] \label{eq15}
\end{align}
\end{figure*}
%%%%%%%%%%%%%%%%%%%%%%%%%%%%%%%%
As observed in \eqref{eq13}, load has impacts on the open-loop plant dynamics of the GFM inverter. Firstly, both $\mathcal{G}_\mathrm{inv}^\mathrm{GFM}(s)$ and $\mathcal{G}_\mathrm{Load}^\mathrm{GFM}(s)$ consist of $L_\mathrm{Load}$ and $R_\mathrm{Load}$ as parameters. that introduce uncertainties in the plant model. Variation of these parameters due to loading of the inverter due to changing generation-load imbalance in the distribution network introduces uncertainties in the plant model \cite{hinf1GFM,hinf2GFM}. Secondly, $i_\mathrm{h}(s)$, imposed by the fundamental and harmonic current load, acts as an exogenous disturbance signal to the plant. Where the latter one is a classical disturbance rejection problem, the former one poses robustness issues. Fig.~\ref{fig:GFLGFMmagphase}\subref{fig:GFM_mag} and Fig.~\ref{fig:GFLGFMmagphase}\subref{fig:GFM_phase} depict the Bode plot of $\mathcal{G}_\mathrm{inv}^\mathrm{GFM}(s)$ that shows the variation in frequency response of the open-loop plant model due to variation in equivalent loading. It is also evident that the uncertain loading variation across the inverter causes significant change in the plant dynamics, especially the effective resonant frequency of the inverter system as shown in Fig.~\ref{fig:GFLGFMmagphase}\subref{fig:GFM_mag}. This deteriorates the transient response of the GFM inverter severely as well as overall system stability \cite{multihinf1,multiloopGFM2}. with the above motivation, this article designs a single-loop $\mu$-synthesis-based stabilizing controller that has robust tracking performance, disturbance rejection and harmonic compensation with improved transient operation under the full range variation in equivalent loading.
% \begin{enumerate}
%     \item Robust stable operation with improved transient operation of the GFM VSI under the full range of load variation, from no load to full load condition.
%     \item Robust tracking performance, disturbance rejection and selective harmonic compensation features of the controller under load uncertainties, from no load to full load condition.
% \end{enumerate}
%%%%%%%%%%%%%%%%%%%%%%%%%END%%%%%%%%%%%%%%%%%%%%%%%%%%%%%%%%%%
%%%%%%%%%%%%%%%%%%%%%%%%%START%%%%%%%%%%%%%%%%%%%%%%%%%%%%%%%
\begin{figure}[t]
	\centering
    \includegraphics[scale=0.175,trim={0cm 0cm 1cm 0cm},clip]{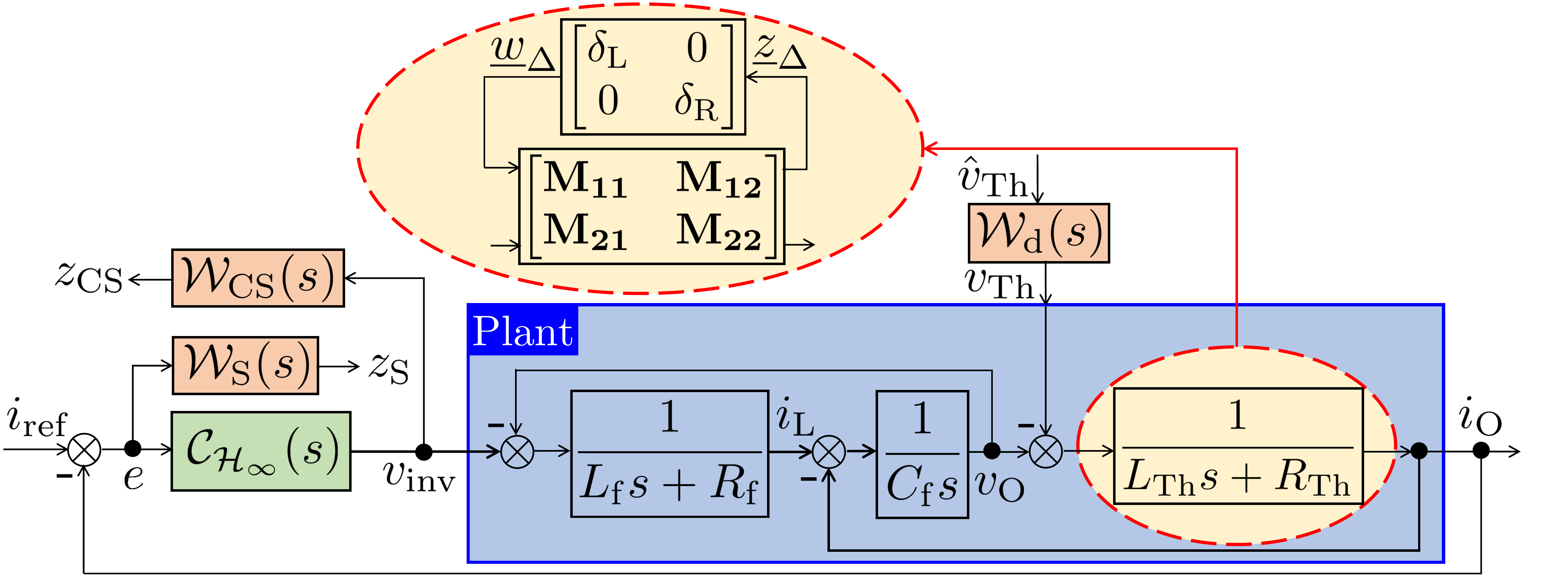}%
	\caption{Proposed $\mu$-synthesis-based robust controller for GFL Inverter.}
	\label{fig:GFLcontrol}
\end{figure}
\begin{figure}[h]
	\centering
    \includegraphics[scale=0.175,trim={0cm 0cm 1cm 0cm},clip]{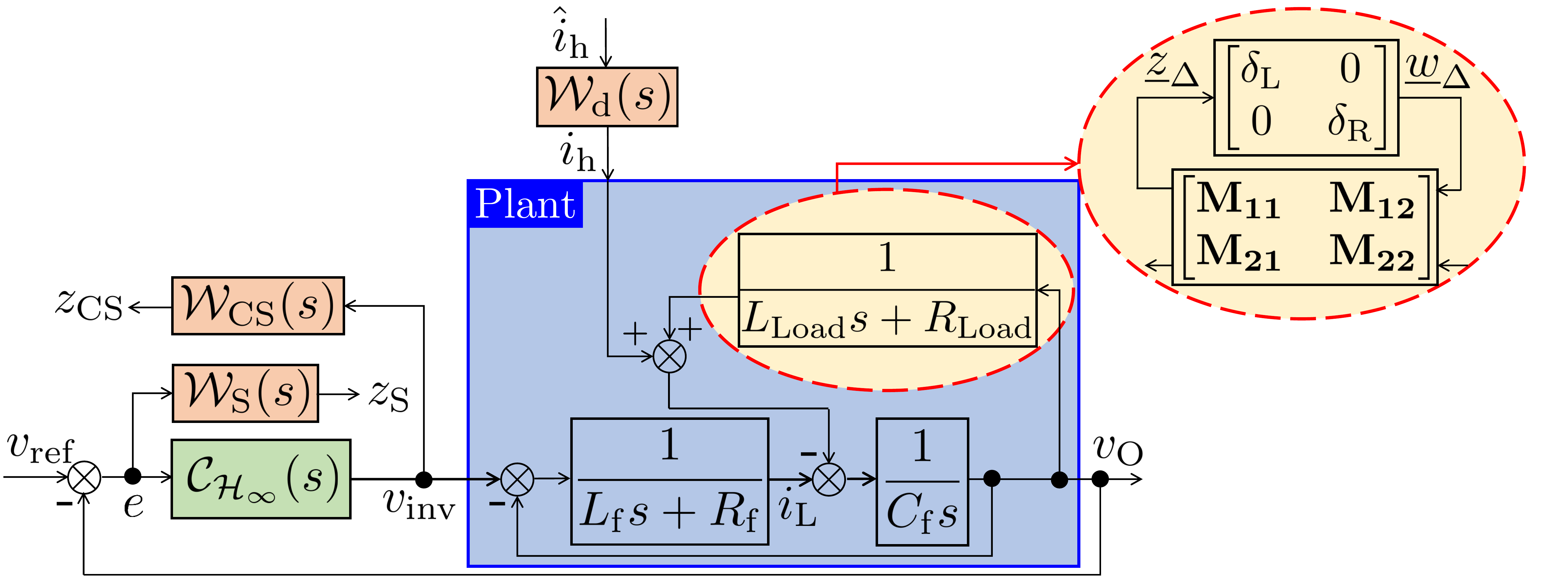}%
	\caption{Proposed $\mu$-synthesis-based robust controller for GFM Inverter.}
	\label{fig:GFMcontrol}
\end{figure}
\section{$\mu$-synthesis-based Generalized Framework for Controller Synthesis}\label{control}
Motivated by Section~\ref{probGFL} and \ref{probGFM}, this section will describe the generalized control framework and required modeling for $\mu$-synthesis-based controller synthesis. 
\subsection{$\mu$-synthesis-based Generalized Framework}
The following observations are common for both GFL and GFM inverter system:\\
$\mathbf{I)}$~The open-loop plant model, $\mathcal{G}_\mathrm{inv}^\mathrm{GFL}(s)$ of \eqref{eq5} and $\mathcal{G}_\mathrm{inv}^\mathrm{GFM}(s)$ of \eqref{eq14} are $3^\mathrm{rd}$ order model with difference numerator type for GFL and GFM inverter respectively. Whereas the disturbance model, $\mathcal{G}_\mathrm{Th}^\mathrm{GFL}(s)$ of \eqref{eq6} and $\mathcal{G}_\mathrm{Load}^\mathrm{GFM}(s)$ of \eqref{eq15}, are $3^\mathrm{rd}$ order models for GFL and GFM respectively. Moreover, there is a voltage-current duality in the plant dynamic model of GFL and GFM inverter. Here for both the models, $v_\mathrm{inv}$ is the \textit{plant~input} signal, $i_\mathrm{O}$, $v_\mathrm{O}$ are the \textit{plant~output} and $v_\mathrm{Th}$, $i_\mathrm{h}$ are the \textit{disturbance} signal for GFL and GFM inverter respectively as given in \eqref{eq4} and \eqref{eq13}.\\
$\mathbf{II)}$~The open-loop plant and disturbance models have similar parametric uncertainties imposed by $L_\mathrm{Th}$, $R_\mathrm{Th}$ and $L_\mathrm{Load}$ and $R_\mathrm{Load}$ for both GFL and GFM inverter respectively. The similarity in evident in Fig.~\ref{fig:GFLcontrol} and Fig.~\ref{fig:GFMcontrol} where the open-loop plant models are shown using block diagram inside the blue boxes and corresponding uncertainty block using dashed ovals for both GFL and GFM inverter respectively.\\
$\mathbf{III)}$~Robust tracking performance, disturbance rejection and harmonic compensation capabilities are desired for controllers of GFL and GFM inverters under the uncertainties.
\par Item $\mathbf{I)}$ and $\mathbf{II)}$ motivate designing a generalized control framework for GFL and GFM inverters based on these similarity and duality property. Third point warrants the $\mu$-synthesis-based robust controller synthesis for addressing multiple objectives. A systematic approach is presented below for designing controller, $\mathcal{C}_{\mathcal{H}_\infty}(s)$ of Fig.~\ref{fig:GFLcontrol} and Fig.~\ref{fig:GFMcontrol} for GFL and GFM inverter respectively using the generalized $\mu$-synthesis-based robust control framework.
\subsection{Modeling of Uncertainty}
It can be observed that in both the GFL and GFM inverter open-loop plant model, the model uncertainty is present as parametric uncertainties in $1^\mathrm{st}$ order transfer function (dashed ovals in Fig.~\ref{fig:GFLcontrol} and Fig.~\ref{fig:GFMcontrol} respectively). A systematic approach is followed here for characterizing and modeling the uncertainly for both the cases as discussed below:
\subsubsection{Characterizing the Uncertainty for GFL Inverter}
Variations in grid impedance (i.e. $L_\mathrm{Th}$ and $R_\mathrm{Th}$) results in real-parametric uncertainties in the GFL plant model. The short-circuit ratio ($\mathrm{SCR}$) is often used to characterize the grid stiffness and can be employed in determining the Thevenin equivalent impedance of the grid at the point-of-connection. $\mathrm{SCR}$ is defined as $(V_\mathrm{PC}^\mathrm{N})^2/\big[S_\mathrm{B}\sqrt{(\omega_\mathrm{N}L_\mathrm{Th})^2+R_\mathrm{Th}^2}\big]$, where $V_\mathrm{PC}^\mathrm{N}$ and $\omega_\mathrm{N}$ are the nominal voltage and frequency at point-of-connection, and $S_\mathrm{B}$ is the rated apparent power of the GFL inverter. Usually the grid at point-of-connection is considered as weak when the $\mathrm{SCR}$ is less than $3$ \cite{ieee1547,loadmodel1}. In this work, with a pre-specified $\mathrm{SCR}$ ($< 3$) and $\mathrm{X}/\mathrm{R}$ ratio ($< 10$), the nominal grid impedance parameters, denoted as $L^\mathrm{Nom}_\mathrm{Th}$ and $R^\mathrm{Nom}_\mathrm{Th}$, are determined. By considering $\pm 100\%$ variations over nominal values, it is assumed that $L_\mathrm{Th}\in [\underline{L}_\mathrm{Th},\bar{L}_\mathrm{Th}]$ and $R_\mathrm{Th}\in [\underline{R}_\mathrm{Th},\bar{R}_\mathrm{Th}]$. It is to be noted that very stiff to extremely weak grid conditions are accommodated with this uncertainty characterization. As a result,
\begin{align}
    L_\mathrm{Th} := L_\mathrm{Th}^\mathrm{Nom} + w_\mathrm{Th}^\mathrm{L}\delta_\mathrm{L},~ R_\mathrm{Th} := R_\mathrm{Th}^\mathrm{Nom} + w_\mathrm{Th}^\mathrm{R}\delta_\mathrm{R},
\end{align}
where, $\delta_\mathrm{L},\delta_\mathrm{R} \in[-1,1]$, $L_\mathrm{Th}^\mathrm{Nom}=\frac{1}{2}[\bar{L}_\mathrm{Th}+\underline{L}_\mathrm{Th}]$, $R_\mathrm{Th}^\mathrm{Nom}=\frac{1}{2}[\bar{R}_\mathrm{Th}+\underline{R}_\mathrm{Th}]$, $w_\mathrm{Th}^\mathrm{L}=\frac{1}{2}[\bar{L}_\mathrm{Th}-\underline{L}_\mathrm{Th}]$, $w_\mathrm{Th}^\mathrm{R}=\frac{1}{2}[\bar{R}_\mathrm{Th}-\underline{R}_\mathrm{Th}]$. 
\subsubsection{Characterizing the Uncertainty for GFM Inverter}
In this case, variation in equivalent loading (i.e. $R_\mathrm{Load}$ and $L_\mathrm{Load}$) results in real-parametric uncertainties in the GFM plant model. In this work, the linear part of the loading is modeled by a series combination of equivalent unknown $R_\mathrm{Load}$ and $L_\mathrm{Load}$ element. These elements are at nominal while the GFM inverter loading is at rated condition. Considering rated VA loading as $S_\mathrm{rated}$, with active power, $P_\mathrm{rated}$, and reactive power, $Q_\mathrm{rated}$, the following holds for nominal values:
\begin{align}
    R_\mathrm{Load}^\mathrm{Nom} = V_\mathrm{N}^2\dfrac{P_\mathrm{rated}}{S^2_\mathrm{rated}}, L_\mathrm{Load}^\mathrm{Nom} = V_\mathrm{N}^2\dfrac{Q_\mathrm{rated}}{S^2_\mathrm{rated}}\dfrac{1}{\omega_\mathrm{N}},
\end{align}
where $V_\mathrm{N}$ and $\omega_\mathrm{N}$ are the nominal voltage and frequency of the network respectively. By considering no-loading and overloading ($200\%$ loading) of GFM, it is assumed that $L_\mathrm{Load}\in [\underline{L}_\mathrm{Load},\bar{L}_\mathrm{Load}]$, $R_\mathrm{Load}\in [\underline{R}_\mathrm{Load},\bar{R}_\mathrm{Load}]$. As a result,
\begin{align}
    L_\mathrm{Load} := L_\mathrm{Load}^\mathrm{Nom} + w_\mathrm{Load}^\mathrm{L}\delta_\mathrm{L},\\ R_\mathrm{Load} := R_\mathrm{Load}^\mathrm{Nom} + w_\mathrm{Load}^\mathrm{R}\delta_\mathrm{R},
\end{align}
where, $\delta_\mathrm{L}, \delta_\mathrm{R} \in[-1,1]$, $L_\mathrm{Load}^\mathrm{Nom}=\frac{1}{2}[\bar{L}_\mathrm{Load}+\underline{L}_\mathrm{Load}]$, $R_\mathrm{Load}^\mathrm{Nom}=\frac{1}{2}[\bar{R}_\mathrm{Load}+\underline{R}_\mathrm{Load}]$, $w_\mathrm{Load}^\mathrm{L}=\frac{1}{2}[\bar{L}_\mathrm{Load}-\underline{L}_\mathrm{Load}]$, and $w_\mathrm{Load}^\mathrm{R}=\frac{1}{2}[\bar{R}_\mathrm{Load}-\underline{R}_\mathrm{Load}]$. 
\subsubsection{Generalized Representation of Uncertainty}
In synthesizing the controller with defined uncertainties in $R \in [\underline{R},\bar{R}]$ and $L \in [\underline{L},\bar{L}]$ when appearing in the form of $1/(Ls+R)$, linear fractional transformation (LFT)\cite{skogestad} can be utilized to convert the model into an upper LFT, $F_\mathrm{U}(\mathbf{M},\mathbf{\Delta})$, given as:
\begin{align}
F_\mathrm{U}(\mathbf{M},\mathbf{\Delta})=\dfrac{1}{sL+R}=\mathrm{M}_{22} + \mathbf{M_{21}\Delta[I-M_{11}\Delta]^{-1}M_{12}},\nonumber\\
\text{with}~\mathbf{M}=\begin{bmatrix}
  \mathbf{M_{11}} & \mathbf{M_{12}} \\
  \mathbf{M_{21}} & \mathrm{M_{22}}
\end{bmatrix},\mathbf{\Delta} = \begin{bmatrix}
  \delta_\mathrm{L} & 0 \\
  0 & \delta_\mathrm{R}
\end{bmatrix},\begin{array}{c}
  \delta_\mathrm{L} \in[-1,1], \\
  \delta_\mathrm{R} \in[-1,1], \\
\end{array}\nonumber\\
\mathbf{M_{11}} =  
  \mathrm{M_{22}}\begin{bmatrix}
     sw^\mathrm{L} & w^\mathrm{R}\\
     sw^\mathrm{L} & w^\mathrm{R}
  \end{bmatrix},
  \begin{array}{c}
  \mathbf{M_{12}} =  
  \mathrm{M_{22}}\begin{bmatrix}
  1 & 1
  \end{bmatrix}^\top, \\
  \mathbf{M_{21}} =  
  \mathrm{M_{22}}\begin{bmatrix}
  sw^\mathrm{L} & w^\mathrm{R}
  \end{bmatrix}, \\
\end{array}\nonumber
\end{align}
where $\mathrm{M_{22}}={-1}/[sL_\mathrm{Nom}+R_\mathrm{Nom}]$,
$w^\mathrm{L}=\frac{1}{2}[\bar{L}-\underline{L}]$, $w^\mathrm{R}=\frac{1}{2}[\bar{L}-\underline{L}]$. It is important to note here that it is a generalized representation of uncertainty for both GFL and GFM inverter where $L=L_\mathrm{Th}$, $R=R_\mathrm{Th}$ for GFL inverter and $L=L_\mathrm{Load}$ $R=R_\mathrm{Load}$ for GFM inverter. 
\begin{figure}[t]
	\centering
    \includegraphics[scale=0.165,trim={0cm 0cm 16cm 0cm},clip]{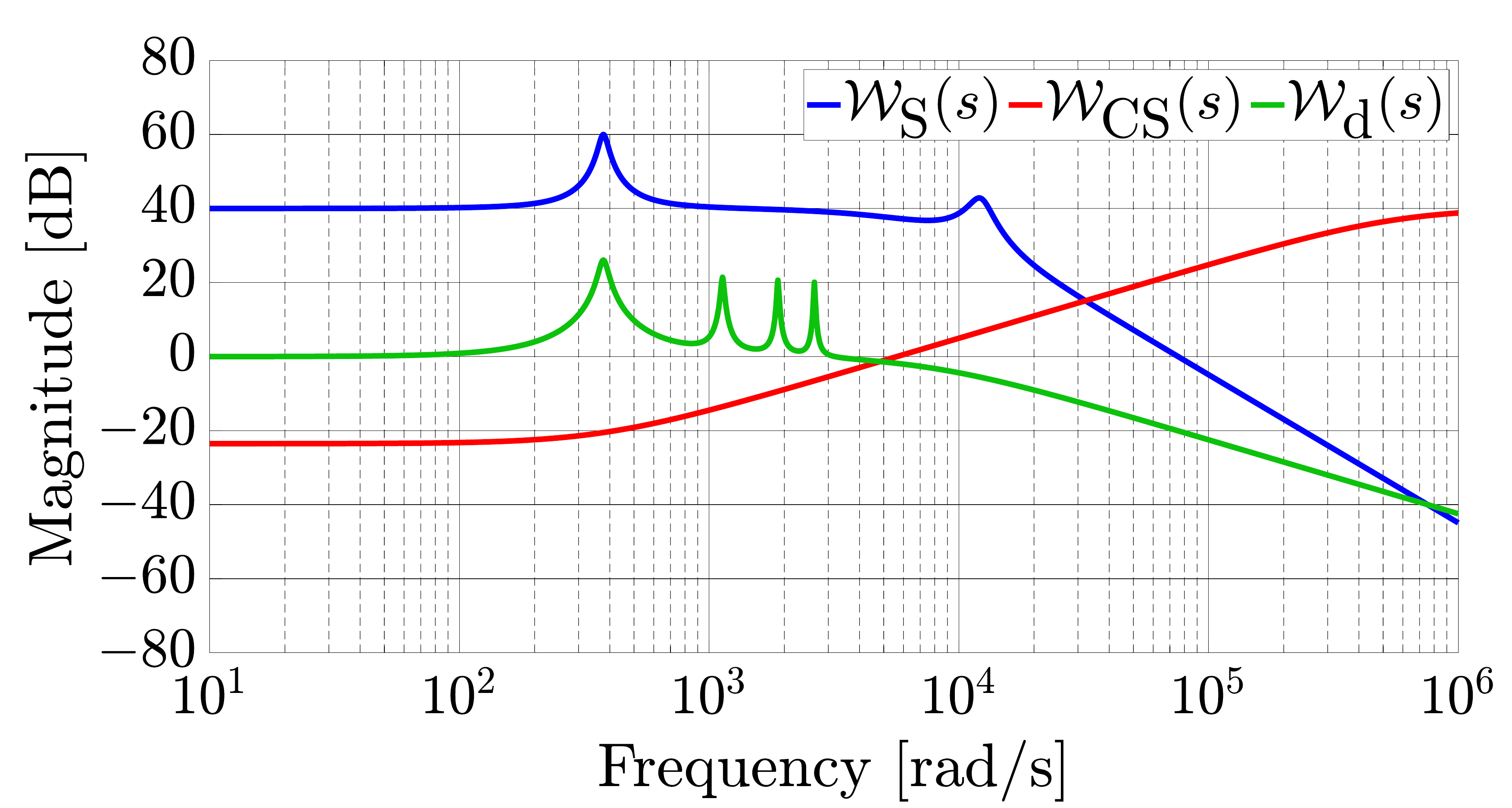}%
	\caption{Bode plots of magnitudes of selected weighting transfer functions.}
	\label{fig:Wbode}
\end{figure}
\subsection{Shaping of Transfer Functions}
The closed-loop objectives in designing the feedback control based on proposed robust controller, $\mathcal{C}_{\mathcal{H}_{\infty}}(s)$, as shown in Fig.~\ref{fig:GFLcontrol} and Fig.~\ref{fig:GFMcontrol} for GFL and GFM inverter, are as follows:\\
$\mathbf{i)~Reference~Tracking:}$~$i_\mathrm{O}$ tracks $i_\mathrm{ref}$ for GFL and $v_\mathrm{O}$ tracks $v_\mathrm{ref}$ for GFM inverter with minimum tracking error.\\
$\mathbf{ii)~Disturbance~Rejection:}$~Effects of $v_\mathrm{Th}$ on $i_\mathrm{O}$ for GFL and effect of $i_\mathrm{h}$ on $v_\mathrm{O}$ for GFM inverter are largely attenuated.\\
$\mathbf{iii)~Control~Effort~Reduction:}$~$v_\mathrm{inv}$ satisfies the respective bandwidth limitations for GFL and GFM inverter. 
\par Based on the objectives, user-defined weighting transfer functions, $\mathcal{W}_\mathrm{S}(s)$, $\mathcal{W}_\mathrm{CS}(s)$, $\mathcal{W}_\mathrm{d}(s)$, are designed. The guidelines for designing the weighting functions are provided below.
\subsubsection{Selection of $\mathcal{W}_\mathrm{S}(s)$}
To shape the sensitivity transfer function, the weighting function, $\mathcal{W}_\mathrm{S}(s)$, is introduced so that
$\mathbf{i)}$~The tracking error, $e$, ($e:=i_\mathrm{ref}-i_\mathrm{O}$ and $e:=v_\mathrm{ref}-v_\mathrm{O}$ for GFL and GFM inverter respectively) at fundamental frequency is small; $\mathbf{ii)}$~Resonance phenomenon of the system is actively damped.\\
$\mathcal{W}_\mathrm{S}(s)$ is modeled to have peaks around $\omega_\mathrm{N}$ and system's resonant frequency, $\omega_\mathrm{r}$ (different in GFL and GFM open-loop plant), with $2^\mathrm{nd}$ order roll-off, $k_\mathrm{S,1}(s)$ and formed as:
\begin{align*}
    \mathcal{W}_\mathrm{S}(s) = k_\mathrm{S,1}(s)\dfrac{s^2+2k_\mathrm{S,2}\zeta\omega_\mathrm{N} s+\omega_\mathrm{N}^2}{s^2+2\zeta\omega_\mathrm{N} s+\omega_\mathrm{N}^2} \dfrac{s^2+2k_\mathrm{S,3}\zeta\omega_\mathrm{r}s+\omega_\mathrm{r}^2}{s^2+2\zeta\omega_\mathrm{r}s+\omega_\mathrm{r}^2},
\end{align*}
where, $k_\mathrm{S,2}$ and $k_\mathrm{S,3}$ are selected to exhibit peaks and $\zeta$ addresses the off-nominal frequency around the nominal values.
\subsubsection{Selection of $\mathcal{W}_\mathrm{CS}(s)$}
$\mathcal{W}_\mathrm{CS}(s)$ is designed to suppress high-frequency control effort to shape the performance of $v_\mathrm{inv}$ for both GFL and GFM controller. Hence, it is designed as a high-pass filter with cut-off frequency at switching frequency for penalizing effect and is ascribed the form:
\begin{align*}
    \mathcal{W}_\mathrm{CS}(s) = k_\mathrm{CS}\dfrac{s+k_\mathrm{CS,1}\omega_\mathrm{N}}{s+k_\mathrm{CS,2}\omega_\mathrm{N}}, ~\text{where}~ k_\mathrm{CS,1}<<k_\mathrm{CS,2}.
\end{align*}
\subsubsection{Selection of $\mathcal{W}_\mathrm{d}(s)$}
$\mathcal{W}_\mathrm{d}(s)$ emphasizes the expected disturbances at fundamental and harmonic frequencies imposed by ${v}_\mathrm{Th}$ and ${i}_\mathrm{h}$ and emphasized by exogenous signal $\hat{v}_\mathrm{Th}$ and $\hat{i}_\mathrm{h}$ for GFL and GFM inverter respectively, as shown in Fig.~\ref{fig:GFLcontrol} and Fig.~\ref{fig:GFMcontrol}. It is designed by a low-pass filter, $k_\mathrm{d}(s)$, with peaks at selected frequencies and is ascribed the form:
\begin{align*}
    \mathcal{W}_\mathrm{d}(s) = k_\mathrm{d}(s)\prod_{\mathrm{h}=1,3,5,7}\dfrac{s^2+2k_\mathrm{d,h}\zeta h\omega_\mathrm{N} s+h^2\omega_\mathrm{N}^2}{s^2+2\zeta h\omega_\mathrm{N} s+h^2\omega_\mathrm{N}^2},
\end{align*}
where, the values of $k_\mathrm{d,h}$ are selected based on the regulated limits of $3^\mathrm{rd}$, $5^\mathrm{th}$, $7^\mathrm{th}$ harmonics in network voltage and current injection with respect to fundamental \cite{ieee2}. A representative of selected weighting functions are shown in Fig.~\ref{fig:Wbode}.
\subsection{Preparing the Generalized Plant}
\begin{figure}[t]
	\centering
	\subfloat[]{\includegraphics[scale=0.09,trim={0cm 0cm 37cm 4.5cm},clip]{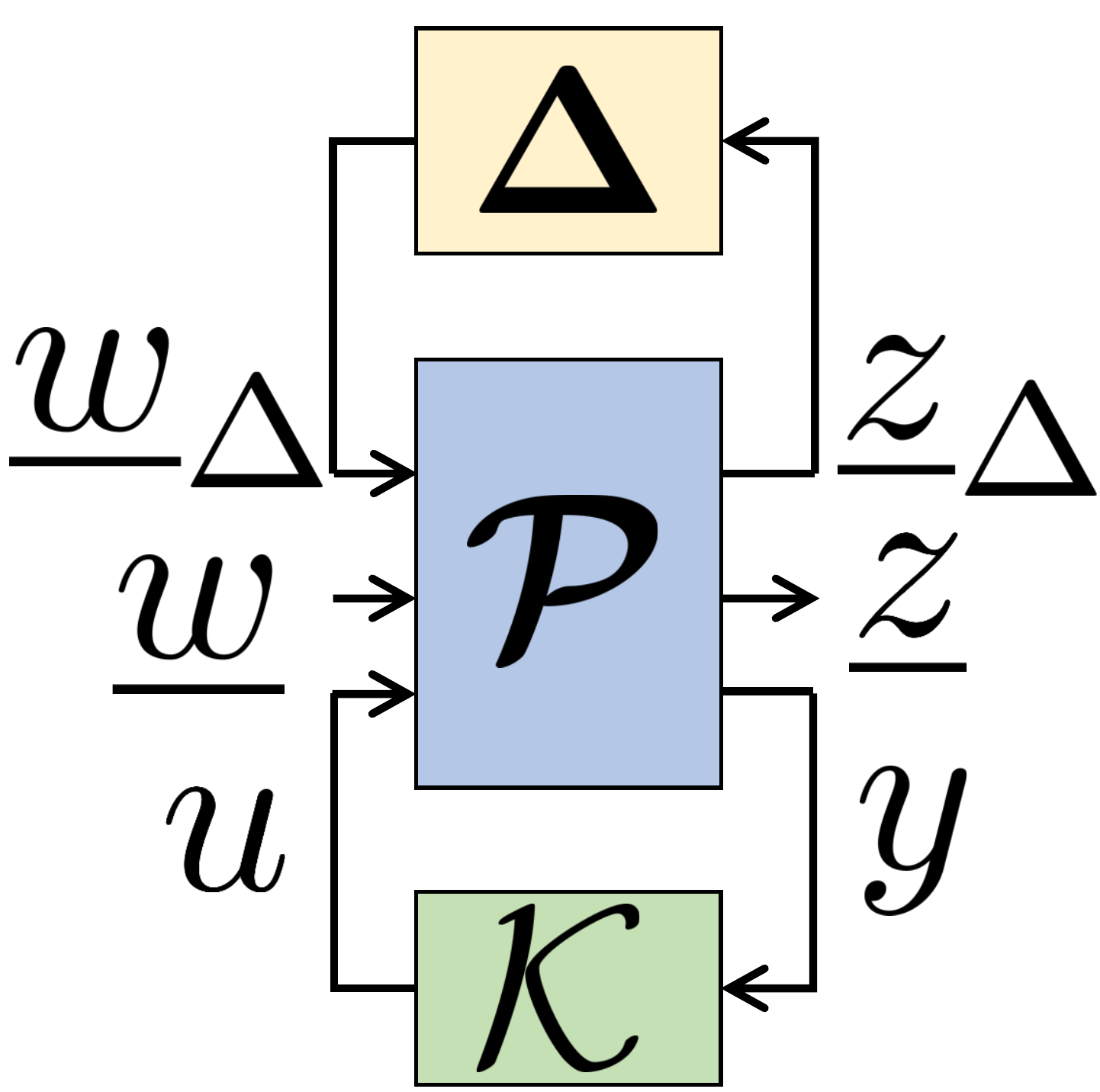}%
	\label{fig:GenCon3}}~
	\subfloat[]{\includegraphics[scale=0.09,trim={0cm 0cm 37cm 5cm},clip]{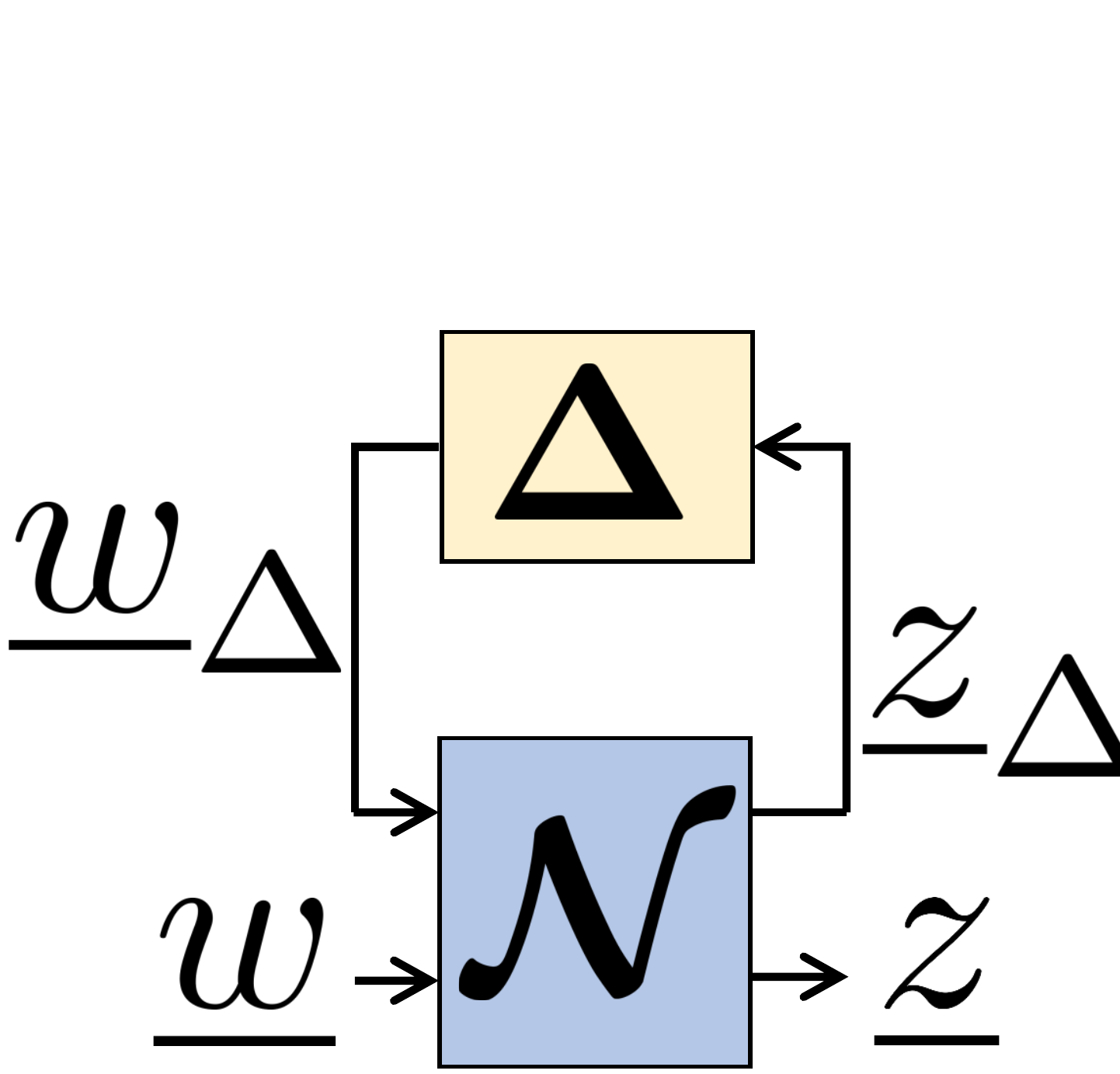}%
	\label{fig:GenCon4}}~
    \subfloat[]{\includegraphics[scale=0.21,trim={0cm 0cm 38cm 12cm},clip]{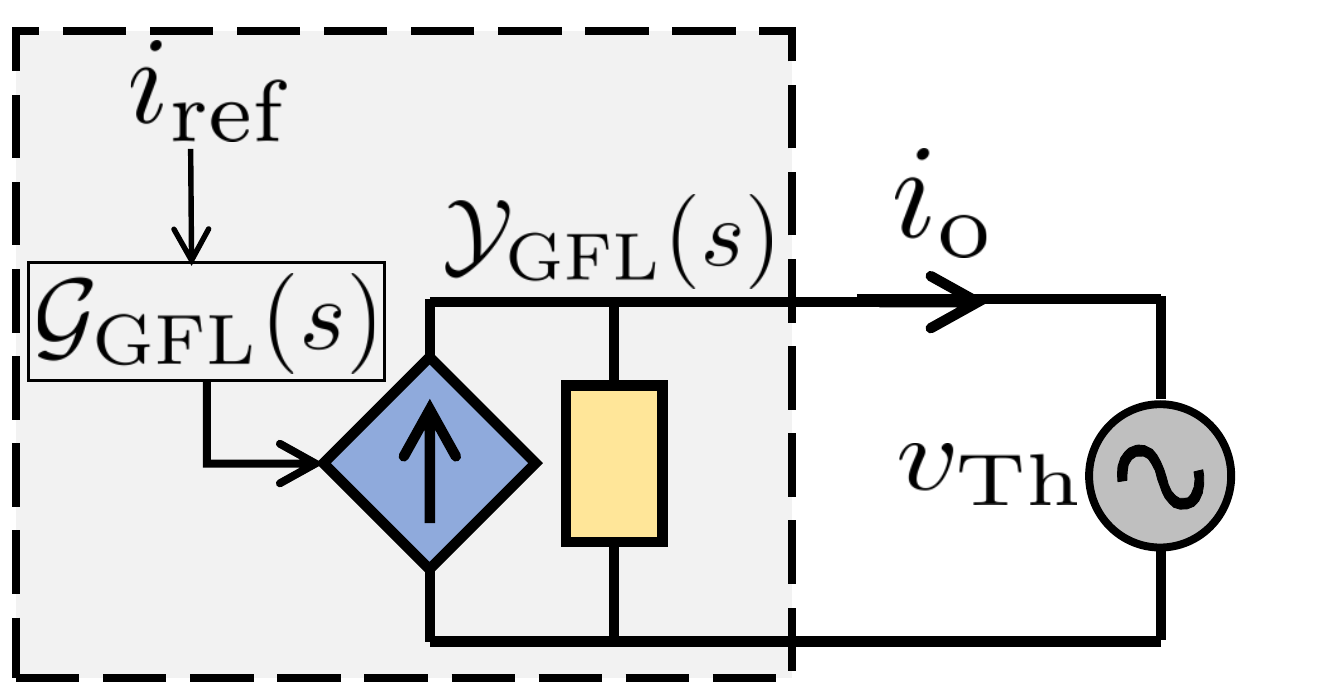}%
	\label{fig:GFL_CL}}~
	\subfloat[]{\includegraphics[scale=0.21,trim={0cm 0cm 38cm 12cm},clip]{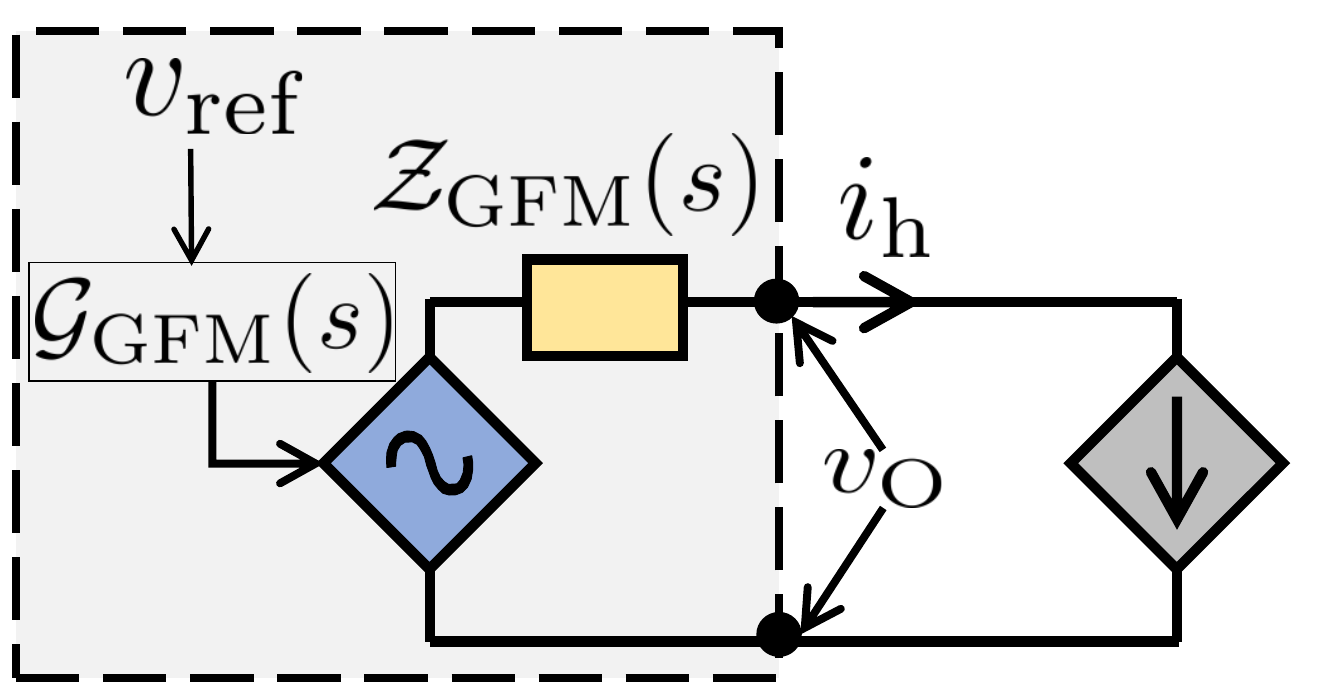}%
	\label{fig:GFM_CL}}
	\caption{Generalized (a) $\Delta$-$\mathcal{P}$-$\mathcal{K}$ control configuration, (b) $\Delta$-$\mathcal{N}$ control configuration, of Fig.~\ref{fig:GFLcontrol} or Fig.~\ref{fig:GFMcontrol} and closed-loop equivalent circuit model of (c) grid-following inverter system, (d) grid-forming inverter system.}
	\label{fig:GenCon}
\end{figure}
In preparation for robust controller design, $\mathcal{C}_{\mathcal{H}_\infty}(s)$, the multi-loop closed-loop block diagram in Fig.~\ref{fig:GFLcontrol} and Fig.~\ref{fig:GFMcontrol} for GFL and GFM inverter respectively are consolidated into the general control configuration in Fig.~\ref{fig:GenCon}\subref{fig:GenCon3} \cite{skogestad}. Here, $\boldsymbol{\mathcal{P}}(s)$ is the generalized multi-input-multi-output (MIMO) plant, $\mathcal{K}(s)$ is the proposed $\mathcal{C}_{\mathcal{H}_\infty}(s)$ controllers to be designed for GFL and GFM inverter and $\mathbf{\Delta}$ is the structured uncertainty. $\underline{w}$ is a vector of the exogenous inputs (e.g., reference, disturbance), $\underline{z}$ are the exogenous outputs (e.g., signals to be regulated). $y$ and $u$ are the controller input and output signals respectively. $\underline{z_\Delta}$ and $\underline{w_\Delta}$ are the vector of input and output signals of structured uncertainty block. Note that in this continuous-time modeling framework, all variables are functions of the Laplace variable, $s$; not explicitly shown for notational convenience. As a result, the generalized MIMO plant maps $\begin{bmatrix}\underline{w_\Delta} & \underline{w} & u\end{bmatrix}^\top$ to $\begin{bmatrix}\underline{z_\Delta} & \underline{z} & y\end{bmatrix}^\top$ as follows:
\begin{align}\label{genconmaps}
    \begin{bmatrix}
    \underline{z_\Delta} \\ \underline{z} \\ y
    \end{bmatrix}=
    \begin{bmatrix}
    \mathcal{P}_{\Delta\Delta} & \mathcal{P}_{\Delta \mathrm{w}} & \mathcal{P}_{\Delta \mathrm{u}} \\ 
    \mathcal{P}_{\mathrm{z}\Delta} & \mathcal{P}_\mathrm{zw} & \mathcal{P}_\mathrm{zu} \\ \mathcal{P}_{\mathrm{y}\Delta} & \mathcal{P}_\mathrm{yw} & \mathcal{P}_\mathrm{yu}
    \end{bmatrix}
    \begin{bmatrix}
    \underline{w_\Delta} \\ \underline{w} \\ u
    \end{bmatrix},
\end{align}
where, the input and output signals are tabulated in Table~\ref{table:IOMaps} for both GFL and GFM inverter system. The detailed MIMO transfer function models of \eqref{genconmaps} for both GFL and GFM inverter systems are provided in Fig.~\ref{MIMO} where 
\begin{align*}
    \mathcal{A}_\mathrm{GFL}&=\dfrac{1}{[L_\mathrm{f}C_\mathrm{f}]s^2 + [R_\mathrm{f}C_\mathrm{f}]s + 1},\\
    \mathcal{B}_\mathrm{GFL}&= \dfrac{L_\mathrm{f}s+R_\mathrm{f}}{[L_\mathrm{f}C_\mathrm{f}]s^2 + [R_\mathrm{f}C_\mathrm{f}]s + 1},\\
    \mathcal{A}_\mathrm{GFM}&=\dfrac{1}{[L_\mathrm{f}C_\mathrm{f}]s^2 + [R_\mathrm{f}C_\mathrm{f}+L_\mathrm{f}]s + [1+R_\mathrm{f}]},\\
    \mathcal{B}_\mathrm{GFM}&= \dfrac{L_\mathrm{f}s+R_\mathrm{f}}{[L_\mathrm{f}C_\mathrm{f}]s^2 + [R_\mathrm{f}C_\mathrm{f}+L_\mathrm{f}]s + [1+R_\mathrm{f}]}.
\end{align*}
If $\mathbf{\Delta}$ is pulled out, then $\boldsymbol{\mathcal{P}}$ and $\mathcal{K}$ can be clubbed together by a lower LFT to form $\boldsymbol{\mathcal{N}}$ in Fig.~\ref{fig:GenCon}\subref{fig:GenCon4} as follows:
\begin{align} \label{genconmaps2}
    \boldsymbol{\mathcal{N}} &= \begin{bmatrix}\mathcal{P}_{\Delta\Delta} & \mathcal{P}_{\Delta \mathrm{w}}\\\mathcal{P}_{\mathrm{z}\Delta} & \mathcal{P}_\mathrm{zw}\end{bmatrix}+\dfrac{\mathcal{K}}{[1-\mathcal{P}_\mathrm{yu}\mathcal{K}]^{-1}}\begin{bmatrix}\mathcal{P}_{\Delta \mathrm{u}}\\\mathcal{P}_\mathrm{zu}\end{bmatrix}\begin{bmatrix}\mathcal{P}_{\mathrm{y}\Delta}\\\mathcal{P}_\mathrm{yw}\end{bmatrix}^\top \nonumber \\
    &= \begin{bmatrix}\mathcal{N}_{\Delta\Delta}&\mathcal{N}_{\Delta \mathrm{w}}\\\mathcal{N}_{\mathrm{z}\Delta}&\mathcal{N}_\mathrm{zw}\end{bmatrix},~\text{where} \begin{bmatrix}\underline{z_\Delta}\\\underline{z}\end{bmatrix}=\boldsymbol{\mathcal{N}}\begin{bmatrix}\underline{w_\Delta}\\\underline{w}\end{bmatrix}.
\end{align}
Therefore, the uncertainty closed-loop transfer function from $\underline{w}$ to $\underline{z}$, $\underline{z}=\boldsymbol{\mathcal{M}}\underline{w}$, is related to $\boldsymbol{\mathcal{N}}$ and $\mathbf{\Delta}$ by an upper LFT where  $\boldsymbol{\mathcal{M}}=\mathcal{N}_\mathrm{zw} + \mathcal{N}_{\mathrm{z}\Delta}\mathbf{\Delta}(\mathbf{I}-\mathcal{N}_{\Delta\Delta}\mathbf{\Delta})^{-1}\mathcal{N}_{\Delta\mathrm{w}}$.
\renewcommand{\arraystretch}{1.2}
\begin{table}[t]
\centering
\caption{INPUT-OUTPUT MAPPING OF THE GENERALIZED PLANT}
\label{table:IOMaps}
\begin{tabular}{c|c|c}
\hline 
$\mathbf{Signals}$ & $\mathbf{GFL~Inverter}$ & $\mathbf{GFM~Inverter}$ \\ \hline 
$\underline{w}$ & $[i_\mathrm{ref}~\hat{v}_\mathrm{Th}]^\top$ $\mathrm{of~Fig.}$~\ref{fig:GFLcontrol} & $[v_\mathrm{ref}~\hat{i}_\mathrm{h}]^\top$ $\mathrm{of~Fig.}$~\ref{fig:GFLcontrol}  \\ \hline 
$\underline{z}$ & $[z_\mathrm{S}~z_\mathrm{CS}]^\top$ $\mathrm{of~Fig.}$~\ref{fig:GFLcontrol} & $[z_\mathrm{S}~z_\mathrm{CS}]^\top$ $\mathrm{of~Fig.}$~\ref{fig:GFMcontrol} \\ \hline 
$u$ & $v_\mathrm{inv}$ $\mathrm{of~Fig.}$~\ref{fig:GFLcontrol} & $v_\mathrm{inv}$ $\mathrm{of~Fig.}$~\ref{fig:GFMcontrol} \\ \hline 
$y$ & $e$ $\mathrm{of~Fig.}$~\ref{fig:GFLcontrol} & $e$ $\mathrm{of~Fig.}$~\ref{fig:GFMcontrol}  \\ \hline 
$\underline{w_\Delta}$ & $\underline{w_\Delta}$ $\mathrm{of~Fig.}$~\ref{fig:GFLcontrol} & $\underline{w_\Delta}$ $\mathrm{of~Fig.}$~\ref{fig:GFMcontrol} \\ \hline 
$\underline{z_\Delta}$ & $\underline{z_\Delta}$ $\mathrm{of~Fig.}$~\ref{fig:GFLcontrol} & $\underline{z_\Delta}$ $\mathrm{of~Fig.}$~\ref{fig:GFMcontrol} \\ \hline 
\end{tabular}
\end{table}
\begin{figure*}[t]
\centering
\scriptsize
\renewcommand{\arraystretch}{1.5}
\begin{tabular}{c|ccc}
\hline
\begin{tabular}[c]{@{}c@{}}
$\mathbf{Type}$
\end{tabular} & \multicolumn{3}{c}{$\boldsymbol{\mathcal{P}}(s)~[\mathcal{P}_{\Delta\Delta} \in \mathbb{C}^{2\times2},~\mathcal{P}_{\Delta \mathrm{w}} \in \mathbb{C}^{2\times2},~\mathcal{P}_{\Delta \mathrm{u}} \in \mathbb{C}^{2\times1},~\mathcal{P}_{\mathrm{z}\Delta} \in \mathbb{C}^{2\times2},~\mathcal{P}_{\mathrm{zw}} \in \mathbb{C}^{2\times2},~\mathcal{P}_{\mathrm{zu}} \in \mathbb{C}^{2\times1},~\mathcal{P}_{\mathrm{y}\Delta} \in \mathbb{C}^{1\times2},~\mathcal{P}_{\mathrm{yw}} \in \mathbb{C}^{1\times2},~\mathcal{P}_{\mathrm{yu}} \in \mathbb{C}]$} \\ \hline
\multirow{3}{*}{$\mathbf{GFL}$} 
& \multicolumn{1}{c|}{$\mathcal{P}_{\Delta\Delta}=\mathbf{M_{11}}-\dfrac{\mathcal{B}_\mathrm{GFL}}{1+\mathrm{M_{22}}\mathcal{B}_\mathrm{GFL}}\mathbf{M_{12}M_{21}}$} 
& \multicolumn{1}{c|}{$\mathcal{P}_{\Delta \mathrm{w}}=\begin{bmatrix} \mathbf{0}_{2\times1} & \dfrac{\mathcal{B}_\mathrm{GFL}\mathcal{W}_\mathrm{d}\mathrm{M}_{22}}{1+\mathrm{M_{22}}\mathcal{B}_\mathrm{GFL}}\mathbf{M_{12}} \end{bmatrix}$} & $\mathcal{P}_{\Delta \mathrm{u}}=\bigg(\mathcal{A}_\mathrm{GFL}-\dfrac{\mathcal{B}_\mathrm{GFL}\mathcal{A}_\mathrm{GFL}\mathrm{M}_{22}}{1+\mathrm{M_{22}}\mathcal{B}_\mathrm{GFL}}\mathbf{M_{12}}\bigg)\mathbf{M_{12}}$
\\ \cline{2-4} 
& \multicolumn{1}{c|}{$\mathcal{P}_{\mathrm{z}\Delta}=\begin{bmatrix}
-\dfrac{\mathcal{W}_\mathrm{S}}{1+\mathrm{M_{22}}\mathcal{B}_\mathrm{GFL}}\mathbf{M_{21}}\\ \mathbf{0}_{1\times2} \end{bmatrix}$} 
& \multicolumn{1}{c|}{$\mathcal{P}_\mathrm{zw}=\begin{bmatrix}
\mathcal{W}_\mathrm{S} & \dfrac{\mathcal{W}_\mathrm{S}\mathcal{W}_\mathrm{d}\mathrm{M_{22}}}{1+\mathrm{M_{22}}\mathcal{B}_\mathrm{GFL}}\\ 0 & 0\end{bmatrix}$} & $\mathcal{P}_\mathrm{zu}=\begin{bmatrix}
-\dfrac{\mathcal{W}_\mathrm{S}\mathrm{M_{22}}\mathcal{A}_\mathrm{GFL}}{1+\mathrm{M_{22}}\mathcal{B}_\mathrm{GFL}}\\ \mathcal{W}_\mathrm{CS} \end{bmatrix}$  
\\ \cline{2-4} 
& \multicolumn{1}{c|}{$\mathcal{P}_{\mathrm{y}\Delta}=-\dfrac{1}{1+\mathrm{M_{22}}\mathcal{B}_\mathrm{GFL}}\mathbf{M_{21}}$} 
& \multicolumn{1}{c|}{$\mathcal{P}_\mathrm{yw}=\begin{bmatrix} 1 & \dfrac{\mathrm{M_{22}\mathcal{W}_\mathrm{d}}}{1+\mathrm{M_{22}}\mathcal{B}_\mathrm{GFL}} \end{bmatrix}$} & $\mathcal{P}_\mathrm{yu}=-\dfrac{\mathrm{M_{2}}\mathcal{A}_\mathrm{GFL}}{1+\mathrm{M_{22}}\mathcal{B}_\mathrm{GFL}}$
\\ \hline
\multirow{3}{*}{$\mathbf{GFM}$} 
& \multicolumn{1}{c|}{$\mathcal{P}_{\Delta\Delta}=\mathbf{M_{11}}-\mathcal{B}_\mathrm{GFM}\mathbf{M_{12}}\mathbf{M_{21}}$} 
& \multicolumn{1}{c|}{$\mathcal{P}_{\Delta\mathrm{w}}=\begin{bmatrix}\mathbf{0}_{2\times1} & -\mathcal{W}_\mathrm{d}\mathcal{B}_\mathrm{GFM}\mathbf{M_{12}}\end{bmatrix}$} & $\mathcal{P}_{\Delta\mathrm{u}}=\mathcal{A}_\mathrm{GFM}\mathbf{M_{12}}$  
\\ \cline{2-4} 
& \multicolumn{1}{c|}{$\mathcal{P}_{\mathrm{z}\Delta}=\begin{bmatrix}-\mathcal{W}_\mathrm{S}\mathcal{B}_\mathrm{GFM}\mathbf{M_{21}}\\ \mathbf{0}_{1\times2} \end{bmatrix}$} 
& \multicolumn{1}{c|}{$\mathcal{P}_{\mathrm{zw}}=\begin{bmatrix} \mathcal{W}_\mathrm{S} & -\mathcal{W}_\mathrm{S}\mathcal{W}_\mathrm{d}\mathcal{B}_\mathrm{GFM} \\ 0 & 0 \end{bmatrix}$} & $\mathcal{P}_{\mathrm{zu}}=\begin{bmatrix} -\mathcal{W}_\mathrm{S}\mathcal{A}_\mathrm{GFM}\\ \mathcal{W}_\mathrm{CS} \end{bmatrix}$
\\ \cline{2-4} 
& \multicolumn{1}{c|}{$\mathcal{P}_{\mathrm{y}\Delta}=\mathcal{B}_\mathrm{GFM}\mathbf{M_{21}}$} 
& \multicolumn{1}{c|}{$\mathcal{P}_{\mathrm{yw}}=\begin{bmatrix} 1 & \mathcal{W}_\mathrm{d}\mathcal{B}_\mathrm{GFM} \end{bmatrix}$} & $\mathcal{P}_{\mathrm{yu}}=-\mathcal{A}_\mathrm{GFM}$ 
\\ \hline
\end{tabular}
\caption{The generalized MIMO transfer function models, $\boldsymbol{\mathcal{P}}(s)$, of \eqref{genconmaps} for both GFL inverter control of Fig.~\ref{fig:GFLcontrol} and GFM inverter control of Fig.~\ref{fig:GFMcontrol}.}
\label{MIMO}
\end{figure*}
%%%%%%%%%%%%%%%%%%%%%%%%%END%%%%%%%%%%%%%%%%%%%%%%%%%%%%%%%%%%
%%%%%%%%%%%%%%%%%%%%%%%%%START%%%%%%%%%%%%%%%%%%%%%%%%%%%%%%%
\section{Controller Synthesis and Stability Analysis}\label{controller}
With reference to the general control configuration of Fig.~\ref{fig:GenCon}\subref{fig:GenCon3}, the standard $\mu$-synthesis-based optimal control problem is to find all stabilizing controllers $\mathcal{K}(s)$ by solving
\begin{align}\label{hinf}
    \min_{\mathcal{K}(s)~\mathrm{stabilizing}}||\boldsymbol{\mathcal{N}}||_\infty,
\end{align}
where, $||.||_{\infty}$ refers to the $\mu$-synthesis norm. This problem can be readily solved using the MATLAB Robust Control Toolbox. An algorithm for solving \eqref{hinf} along with the theoretical underpinnings of this optimization problem can be found in \cite{skogestad}. Upon finding a stabilizing controller, the requirement of stability and performance of the closed-loop system are needed to be checked and can be summarized as follows:
\begin{align}
    &\mathrm{Nominal~Stable(NS)}: \boldsymbol{\mathcal{N}}~\mathrm{is~internally~stable},\label{RP1}\\
    &\mathrm{Nominal~Performance(NP)}: ||\mathcal{N}_\mathrm{zw}||_\infty < 1~\mathrm{\&~NS},\label{RP2}\\
    &\mathrm{Robust~Stable(RS)}: \mu_\Delta(\mathcal{N}_{\Delta\Delta}) < 1,\forall \omega,~\mathrm{\&~NS},\label{RP3}\\
    &\mathrm{Robust~Performance(RP)}: \mu_{\bar{\Delta}}(\boldsymbol{\mathcal{N}}) < 1,\forall \omega,~\mathrm{\&~NS},\label{RP4}
\end{align}
where, $\mu_\Delta(\mathcal{N}_{\Delta\Delta})$ and $\mu_{\bar{\Delta}}(\boldsymbol{\mathcal{N}})$ are the structured singular values of $\mathcal{N}_{\Delta\Delta}$ and $\boldsymbol{\mathcal{N}}$ for the allowed structure of $\boldsymbol{\Delta}$ and $\boldsymbol{\bar{\Delta}}:=\mathrm{diag}(\boldsymbol{\Delta},\boldsymbol{\Delta}_\mathrm{P})$ respectively with $\boldsymbol{\Delta}_\mathrm{P}$ being an unstructured uncertainty \cite{skogestad}. It is necessary to check whether stabilizing controller, $\mathcal{K}(s)$ of \eqref{hinf} satisfies all the conditions of \eqref{RP1}-\eqref{RP4} to analyze robust performance of the controller. In this work, an iterative approach is followed for $\mu$-synthesis problem (i.e. finding the stabilizing controller that minimizes a given $\mu$-condition). The parameters in either performance weights (i.e. $\mathcal{W}_\mathrm{S}(s)$, $\mathcal{W}_\mathrm{CS}(s)$, $\mathcal{W}_\mathrm{d}(s)$) or uncertainty weights (i.e. $w^\mathrm{L}$, $w^\mathrm{R}$ are adjusted and then solved \eqref{hinf} until conditions of \eqref{RP1}-\eqref{RP4} are all satisfied. The optimal controller $\mathcal{K}(s)$ will have an order similar to the order of $\boldsymbol{\mathcal{P}}$. Thus, before implementation in actual inverter control board, model order reduction is used to obtain a lower order controller using  MATLAB’s \textit{balred} command. Moreover, bilinear transformation is used in the discretization stage of the resulting controller.
\renewcommand{\arraystretch}{1.2}
\begin{table}[t]
\centering
\caption{STABILITY ASSESSMENT OF RESULTING CONTROLLER}
\label{table:robust}
\begin{tabular}{c|c|c|c|c}
\hline 
$\mathbf{Controller}$ & $\mathbf{NS}$ & $\mathbf{NP}$ & $\mathbf{RS}$ & $\mathbf{RP}$ \\ \hline 
$\mathrm{GFL~Inverter}$ & $\checkmark$ & $0.49$ & $0.91$ & $0.94$ \\ \hline
$\mathrm{GFM~Inverter}$ & $\checkmark$ & $0.47$ & $0.94$ & $0.98$ \\ \hline
\end{tabular}
\end{table}
\subsubsection{Analysis of Resulting Controller for GFL Inverter} 
\begin{figure}[t]
	\centering
	\subfloat[]{\includegraphics[scale=0.122,trim={0cm 0cm 16cm 1cm},clip]{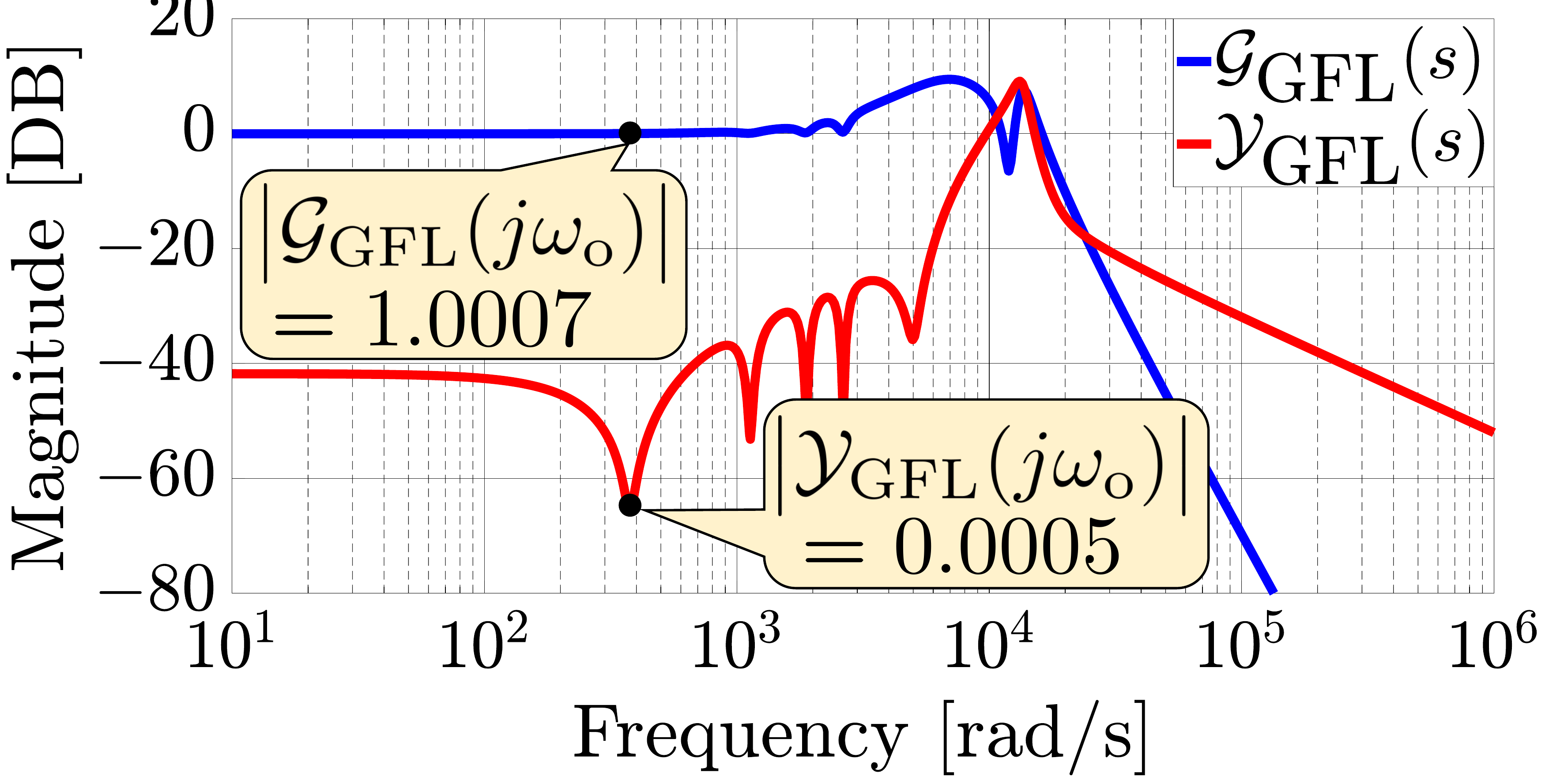}%
	\label{fig:GYMag}}~
	\subfloat[]{\includegraphics[scale=0.122,trim={0cm 0cm 16cm 1cm},clip]{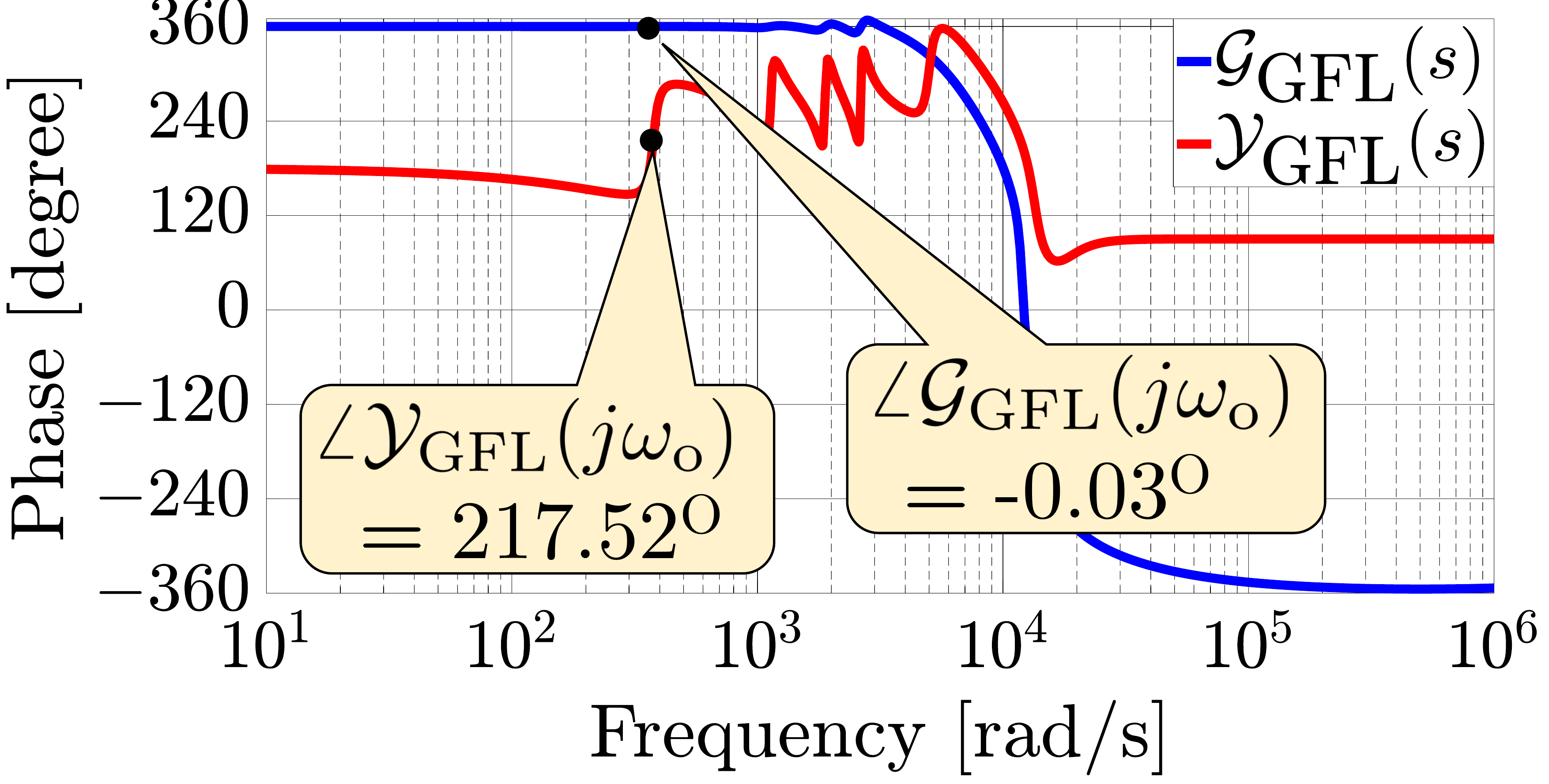}%
	\label{fig:GYPhase}}
	\caption{Bode plots, (a) magnitudes, (b) phase, of $\mathcal{G}_\mathrm{GFL}$, $\mathcal{Y}_\mathrm{GFL}$ of Fig.~\ref{fig:GenCon}(c).}
	\label{fig:GY}
\end{figure}

Following the procedure of synthesizing the optimal controller for GFL inverter, a $13^\mathrm{th}$ order $\mathcal{C}_{\mathcal{H_\infty}}(s)$ of Fig.~\ref{fig:GFLcontrol} is found to perform well. The closed-loop stability and desired performances are met as summarized in Table~\ref{table:robust}. The closed-loop model for GFL inverter with negative feedback loop transfer function with resulting controller, $\mathcal{C}_{\mathcal{H_\infty}}(s)$, in Fig.~\ref{fig:GFLcontrol} can be derived by substituting $v_\mathrm{inv}(s)=\mathcal{C}_{\mathcal{H_\infty}}(s)[i_\mathrm{ref}-i_\mathrm{O}]$ in \eqref{eq4}. It can be written as $i_\mathrm{O} = \mathcal{G}_\mathrm{GFL}(s)i_\mathrm{ref} - \mathcal{Y}_\mathrm{GFL}(s)v_\mathrm{Th}$ and represented as Norton's equivalent model connected to a voltage source as shown in Fig.~\ref{fig:GenCon}\subref{fig:GFL_CL}. For an example, at nominal plant condition with resulting optimal controller, the Bode plot of $\mathcal{G}_\mathrm{GFL}(s)$ and $\mathcal{Y}_\mathrm{GFL}(s)$ are shown in Fig.~\ref{fig:GY}\subref{fig:GYMag} and Fig.~\ref{fig:GY}\subref{fig:GYPhase} respectively. It is observed that $|\mathcal{G}_\mathrm{GFL}(j\omega_\mathrm{N})|$, $\angle \mathcal{G}_\mathrm{GFL}(j\omega_\mathrm{N})$ and $|\mathcal{Y}_\mathrm{GFL}(j\omega_\mathrm{N})|$ are $\approx 1$, $\approx 0^{\circ}$ and $\approx 0$ respectively that leads to $i_\mathrm{O} \approx i_\mathrm{ref}$ at fundamental frequency. 
\subsubsection{Analysis of Resulting Controller for GFM Inverter}
\begin{figure}[t]
	\centering
	\subfloat[]{\includegraphics[scale=0.122,trim={0cm 0cm 16cm 1cm},clip]{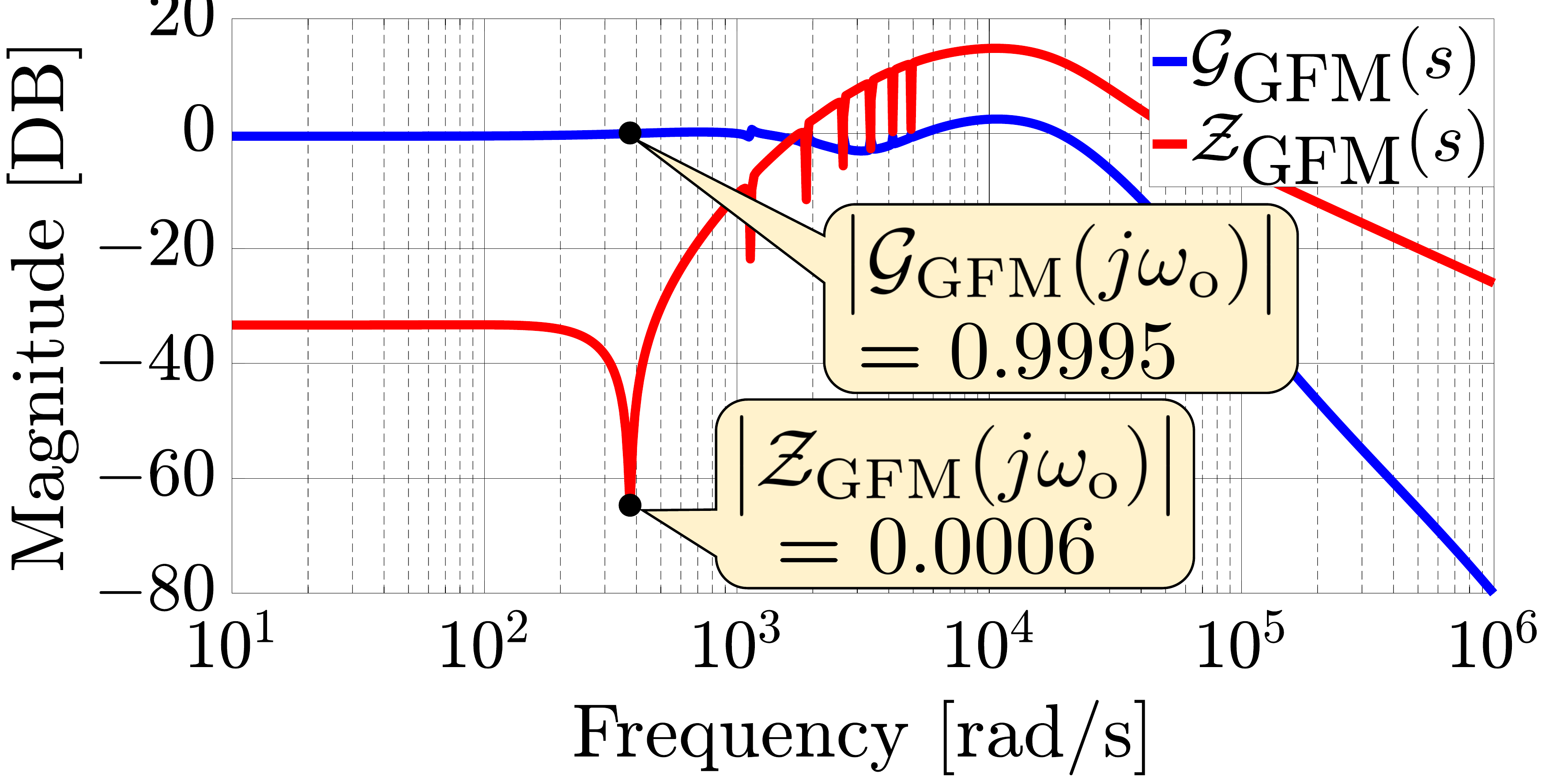}%
	\label{fig:GZMag}}~
	\subfloat[]{\includegraphics[scale=0.122,trim={0cm 0cm 16cm 1cm},clip]{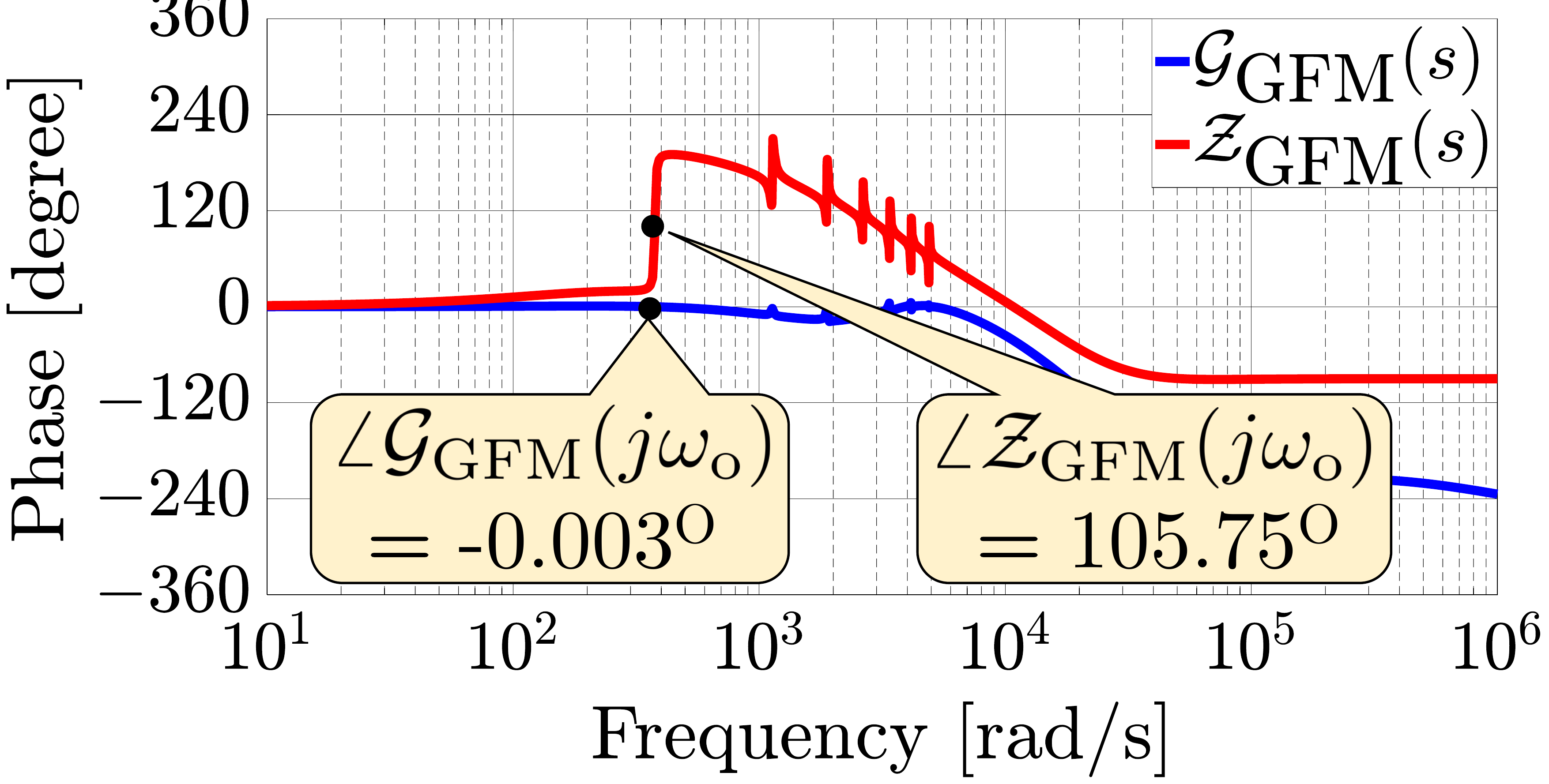}%
	\label{fig:GZPhase}}
	\caption{Bode plots, (a) magnitudes, (b) phase, of $\mathcal{G}_\mathrm{GFM}$, $\mathcal{Z}_\mathrm{GFM}$ of Fig.~\ref{fig:GenCon}(d).}
	\label{fig:GZ}
\end{figure}

Following the procedure of synthesizing the optimal controller for GFM inverter, a $14^\mathrm{th}$ order $\mathcal{C}_{\mathcal{H_\infty}}(s)$ of Fig.~\ref{fig:GFMcontrol} is found to be sufficient and performs well. The closed-loop stability and desired performances are met as summarized in Table~\ref{table:robust}. The closed-loop model for GFM inverter with negative feedback loop transfer function with resulting controller, $\mathcal{C}_{\mathcal{H_\infty}}(s)$, in Fig.~\ref{fig:GFMcontrol} can be derived by substituting $v_\mathrm{inv}(s)=\mathcal{C}_{\mathcal{H_\infty}}(s)[v_\mathrm{ref}-v_\mathrm{O}]$ in \eqref{eq13}. It can be written as $v_\mathrm{O} = \mathcal{G}_\mathrm{GFM}(s)v_\mathrm{ref} - \mathcal{Z}_\mathrm{GFM}(s)i_\mathrm{h}$ and represented as Thevenin's equivalent model connected across a current source as shown in Fig.~\ref{fig:GenCon}\subref{fig:GFM_CL}. For an example, at nominal plant condition with resulting optimal controller, the Bode plot of $\mathcal{G}_\mathrm{GFM}(s)$ and $\mathcal{Z}_\mathrm{GFM}(s)$ are shown in Fig.~\ref{fig:GZ}\subref{fig:GZMag} and Fig.~\ref{fig:GZ}\subref{fig:GZPhase} respectively. It is observed that $|\mathcal{G}_\mathrm{GFM}(j\omega_\mathrm{N})|$, $\angle \mathcal{G}_\mathrm{GFM}(j\omega_\mathrm{N})$ and $|\mathcal{Z}_\mathrm{GFM}(j\omega_\mathrm{N})|$ are $\approx 1$, $\approx 0^{\circ}$ and $\approx 0$ respectively that leads to $v_\mathrm{O} \approx v_\mathrm{ref}$ at fundamental frequency. 
%%%%%%%%%%%%%%%%%%%%%%%%%END%%%%%%%%%%%%%%%%%%%%%%%%%%%%%%%%%%
%%%%%%%%%%%%%%%%%%%%%%%%%START%%%%%%%%%%%%%%%%%%%%%%%%%%%%%%%%%%
\section{Experimental Results and Verification}\label{result}
\begin{figure*}[t]
	\centering
	\subfloat[]{\includegraphics[scale=0.25,trim={0cm 0cm 27cm 2cm},clip]{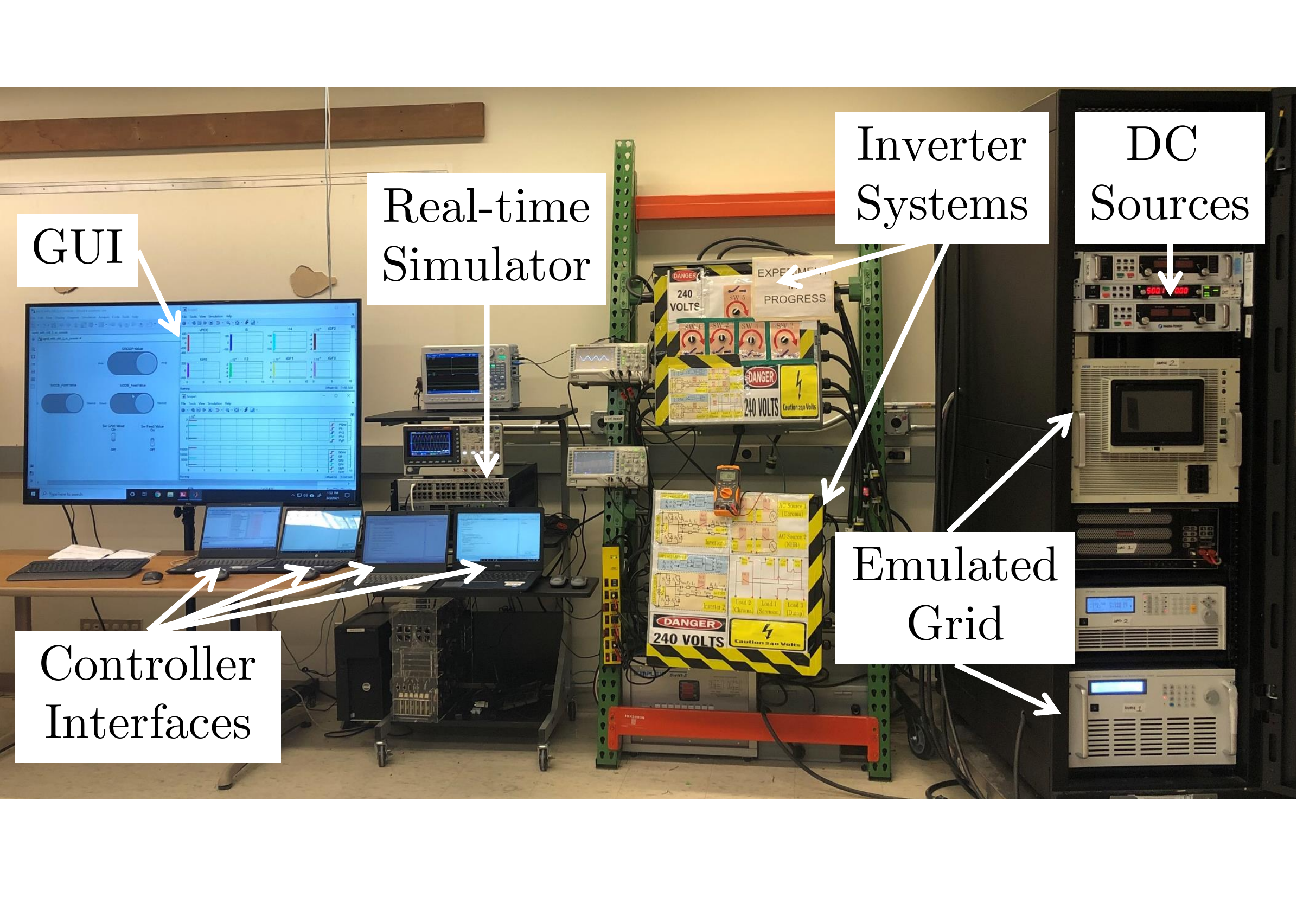}%
	\label{fig:setup}}~
	\subfloat[]{\includegraphics[scale=0.25,trim={2.5cm 0cm 2.5cm 0cm},clip]{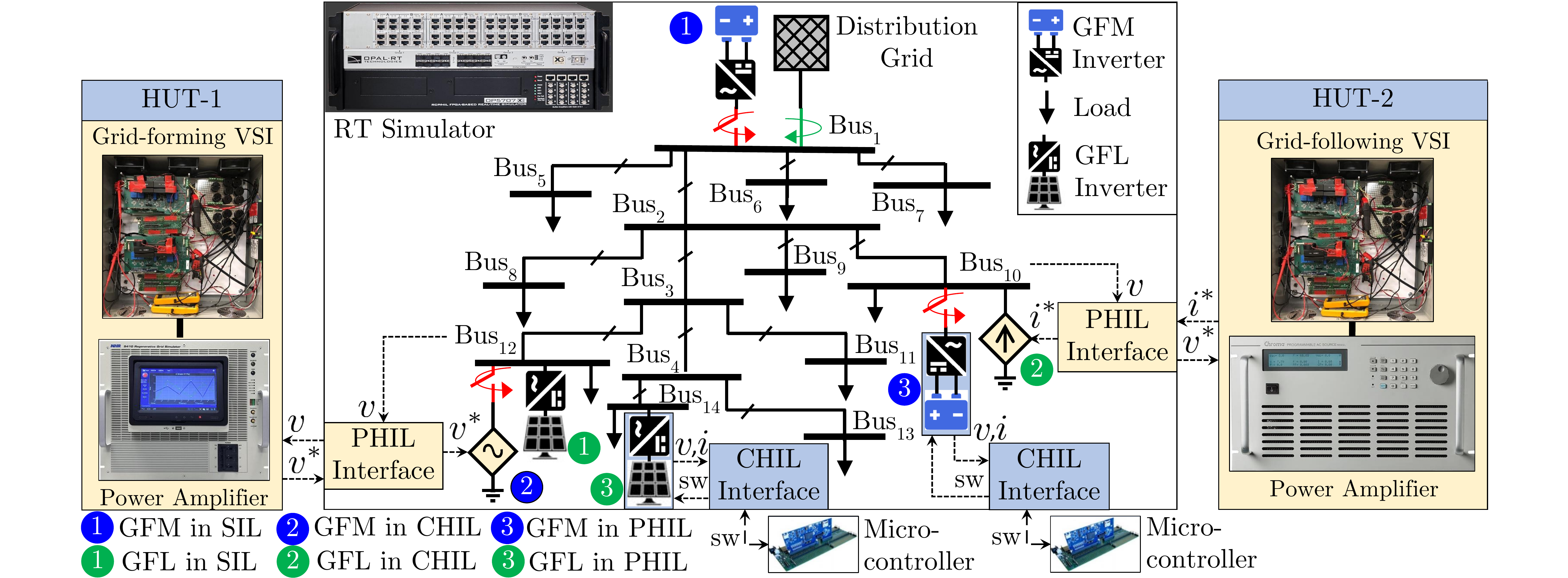}%
	\label{fig:PHILsetup}}
	\caption{(a) The laboratory-based experimental hardware setup, (b) the configuration of SIL-CHIL-PHIL-based experimental setup.}
	\label{fig:hardwaresetup}
\end{figure*}
\renewcommand{\arraystretch}{1.2}
\begin{table}[t]
\centering
\caption{$1$-PHASE INVERTER SYSTEM UNDER STUDY}
\label{table:data}
\begin{tabular}{c|c}
\hline 
$\mathbf{Inverter}$ & $\mathbf{Value}$    \\ \hline 
$\mathrm{Ratings}$ ($1$-$\phi$) & $240~\mathrm{V}$, $60~\mathrm{Hz}$, $1.67~\mathrm{kVA}$, $0.9~\mathrm{pf}$ \\ \hline
$\mathrm{Inverter~Parameters}$ & $V_\mathrm{dc}$ = $500~\mathrm{V}$, $f_\mathrm{sw}$ = $20~\mathrm{kHz}$ \\ \hline
$\mathrm{Filter~Parameters}$ & $L_\mathrm{f}$ = $2~\mathrm{mH}$, $R_\mathrm{f}$ = $0.2$~$\Omega$, $C_\mathrm{f}$ = $20~\mu\mathrm{F}$ \\ \hline
\end{tabular}
\end{table}
\subsection{Experimental Configuration}
A combined system-in-the-loop (SIL), controller hardware-in-the-loop (CHIL) and power hardware-in-the-loop (PHIL) based experimental validation is conducted in order to evaluate the efficacy and viability of the proposed $\mu$-synthesis-based controller for single-phase GFL and GFM inverters. The ratings and parameters of the inverter systems are tabulated in Table~\ref{table:data}. The laboratory-based experimental setup is shown in Fig.~\ref{fig:hardwaresetup}\subref{fig:setup}. The configuration is shown in Fig.~\ref{fig:hardwaresetup}\subref{fig:PHILsetup} and described below:
\subsubsection{Real-time Simulation and SIL Configuration}
A residential sub-network of North American low voltage distribution feeder from CIGRE Task Force C$6.04.02$ \cite{testsystem}, affiliated with CIGRE Study Committee C$6$ is emulated using eMEGASIM platform inside the OP$5700$ RT-simulator (RTS) manufactured by OPAL-RT. The original ratings of load at each bus and line parameters are modified in order to make it compatible with the voltage rating and power capacity of the laboratory. Moreover, the test system is modified by including sufficient non-linear loads at various buses while respecting the recommended limits of harmonic distortions mentioned in \cite{ieee2}. As part of SIL-setup, one GFM and one GFL inverter are emulated entirely (i.e. both power circuit and the control with proposed $\mu$-synthesis-based controller) inside the RTS, connected at $\mathrm{Bus}_{1}$ and $\mathrm{Bus}_{12}$ respectively as shown in Fig.~\ref{fig:hardwaresetup}\subref{fig:PHILsetup}.
\subsubsection{Controller Hardware-in-the-loop Configuration}
As part of CHIL-setup, one GFM and one GFL inverter system are emulated with only power circuit inside the RTS, connected at $\mathrm{Bus}_{10}$ and $\mathrm{Bus}_{14}$ respectively as shown in Fig.~\ref{fig:hardwaresetup}\subref{fig:PHILsetup}. The proposed $\mu$-synthesis-based control logic of both GFL and GFM inverter systems are realized on two Texas-Instruments $\mathrm{TMS}320\mathrm{F}28379\mathrm{D}$, $16$/$12$-bit floating-point $200$~MHz Delfino micro-controller boards interfaced with RTS. 
\subsubsection{Power Hardware-in-the-loop Configuration}
As part of PHIL-setup, one GFM (HUT-$1$ in Fig.~\ref{fig:hardwaresetup}\subref{fig:PHILsetup}) and one GFL inverter (HUT-$2$ in Fig.~\ref{fig:hardwaresetup}\subref{fig:PHILsetup}), connected at $\mathrm{Bus}_{12}$ and $\mathrm{Bus}_{10}$ respectively, are physically realized outside the RTS. The ideal transformer model (ITM) based PHIL interface logic \cite{PHILpaper} is adapted for both the hardware-under-tests (HUTs'). In HUT-$1$ the physical inverter system, fed by MAGNA-POWER programmable DC power supply, is interfaced with low-cost Texas-Instruments TMS$320$F$28379$D, $16$/$12$-bit floating-point $200$~MHz Delfino micro-controller boards employed with proposed $\mu$-synthesis control logic for GFM inverter. The power terminals of the inverter are connected with a power amplifier realized by NHR $9410$ regenerative grid simulator. On the other hand, in HUT-$2$ the physical inverter system, fed by another MAGNA-POWER programmable DC power supply, is interfaced with another low-cost Texas-Instruments TMS$320$F$28379$D, $16$/$12$-bit floating-point $200$~MHz Delfino micro-controller boards employed with proposed $\mu$-synthesis control logic for GFL inverter. The power terminals of the inverter are connected with a power amplifier realized by Chroma $61605$ programmable ac power source.
%%%%%%%%%%%
\begin{figure}[t]
	\centering
	\subfloat[]{\includegraphics[scale=0.11,trim={0cm 0cm 13cm 0cm},clip]{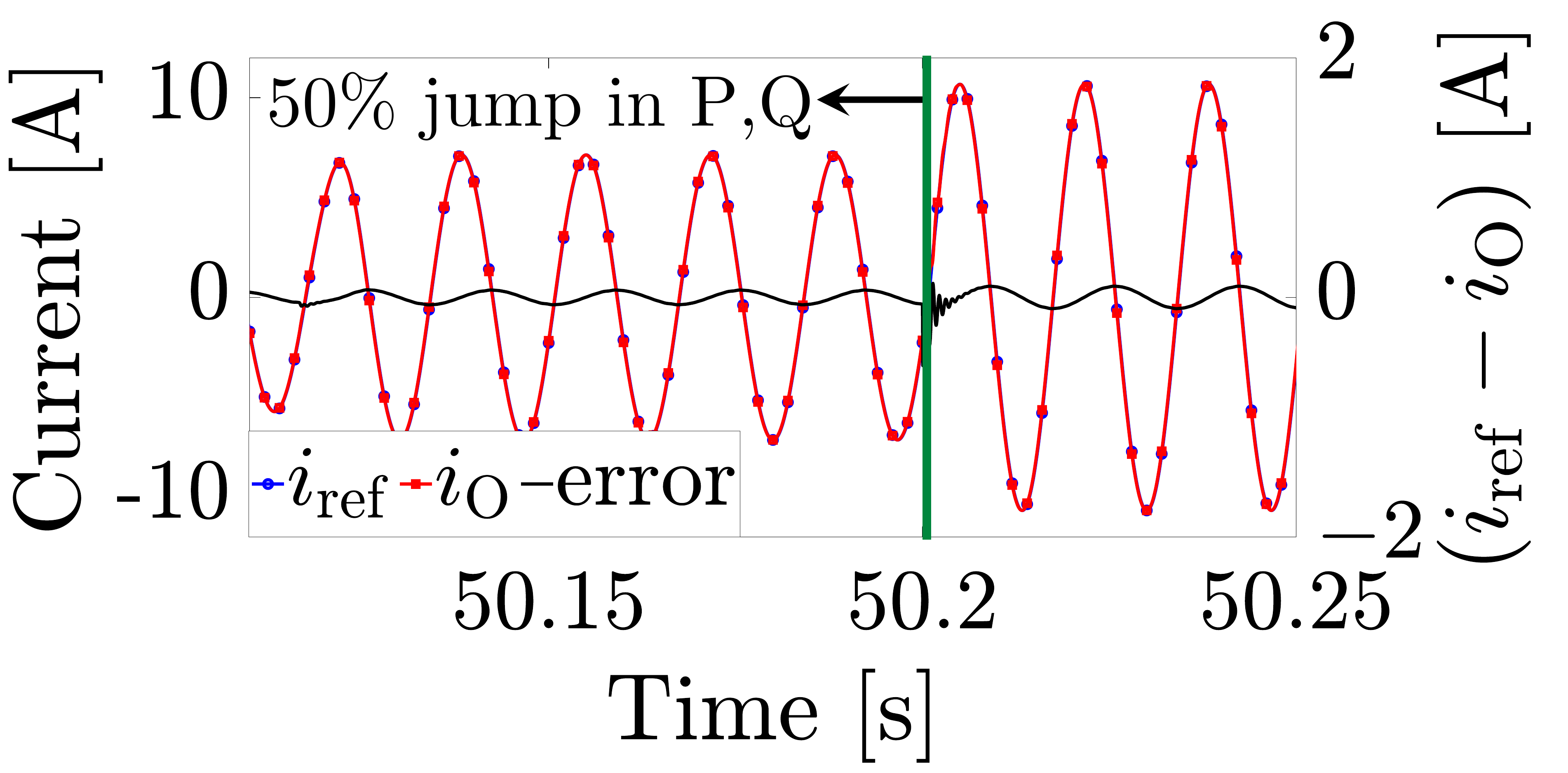}%
	\label{fig:gfl_chil1a}}~
	\subfloat[]{\includegraphics[scale=0.11,trim={0cm 0cm 13cm 0cm},clip]{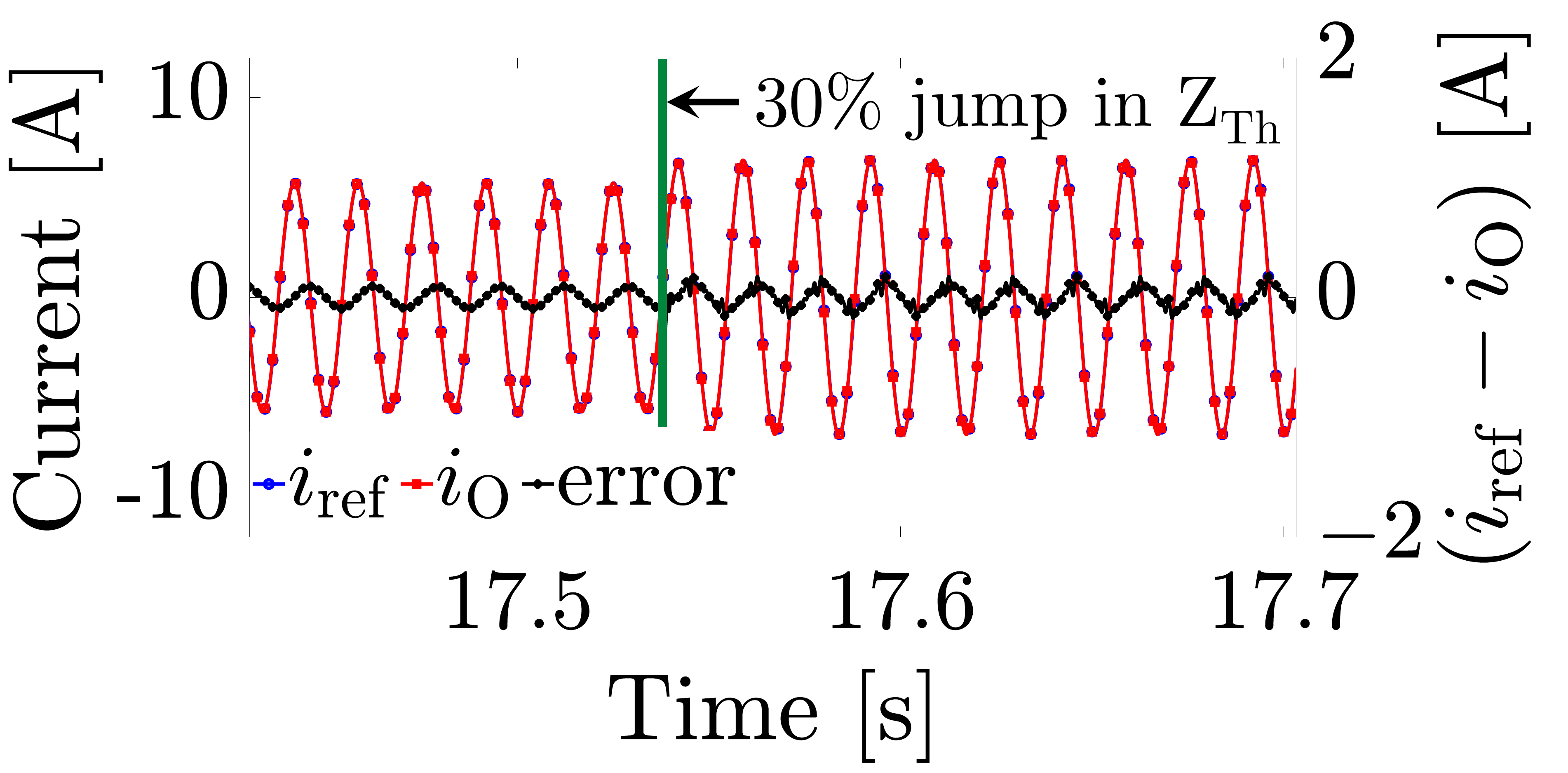}%
	\label{fig:gfl_chil2a}}
	\caption{The current reference tracking capability of GFL inverter with proposed $\mu$-synthesis-based optimal controller in CHIL setup at $\mathrm{Bus}_{14}$ of Fig.~\ref{fig:hardwaresetup}(b) in (a) $\mathtt{CASE}$-$\mathtt{1}$, (b) $\mathtt{CASE}$-$\mathtt{2}$.}
	\label{fig:gfl_chil1}
\end{figure}
\begin{figure}[t]
	\centering
    \includegraphics[scale=0.25,trim={0cm 0cm 20cm 0cm},clip]{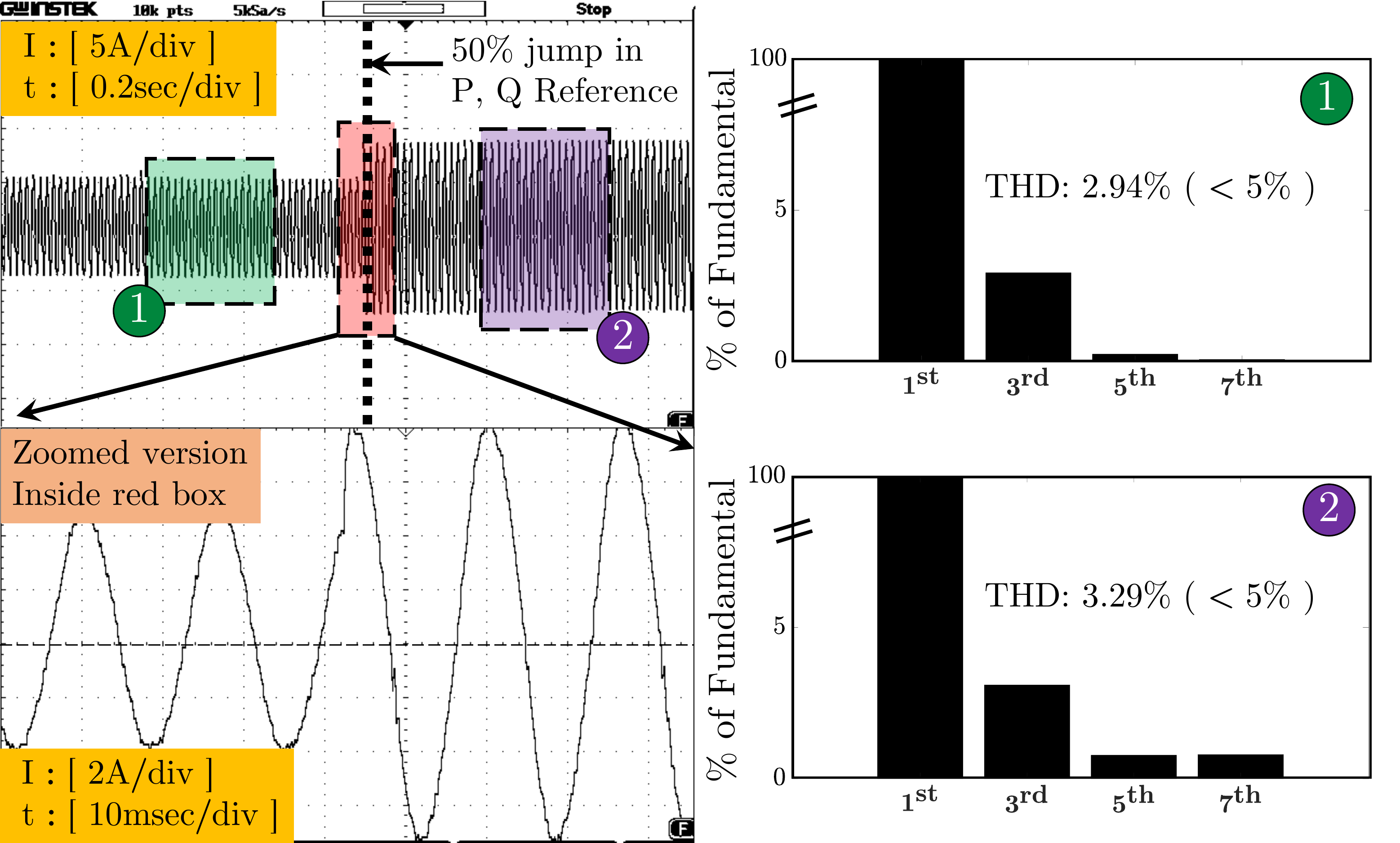}%
	\caption{The current reference tracking capability of GFL inverter in PHIL setup at $\mathrm{Bus}_{10}$ of Fig.~\ref{fig:hardwaresetup}(b) in $\mathtt{CASE}$-$\mathtt{1}$ with the proposed optimal controller.}
	\label{fig:GFLcase1C}
\end{figure}
\begin{figure}[t]
	\centering
    \includegraphics[scale=0.25,trim={0cm 0cm 20cm 0cm},clip]{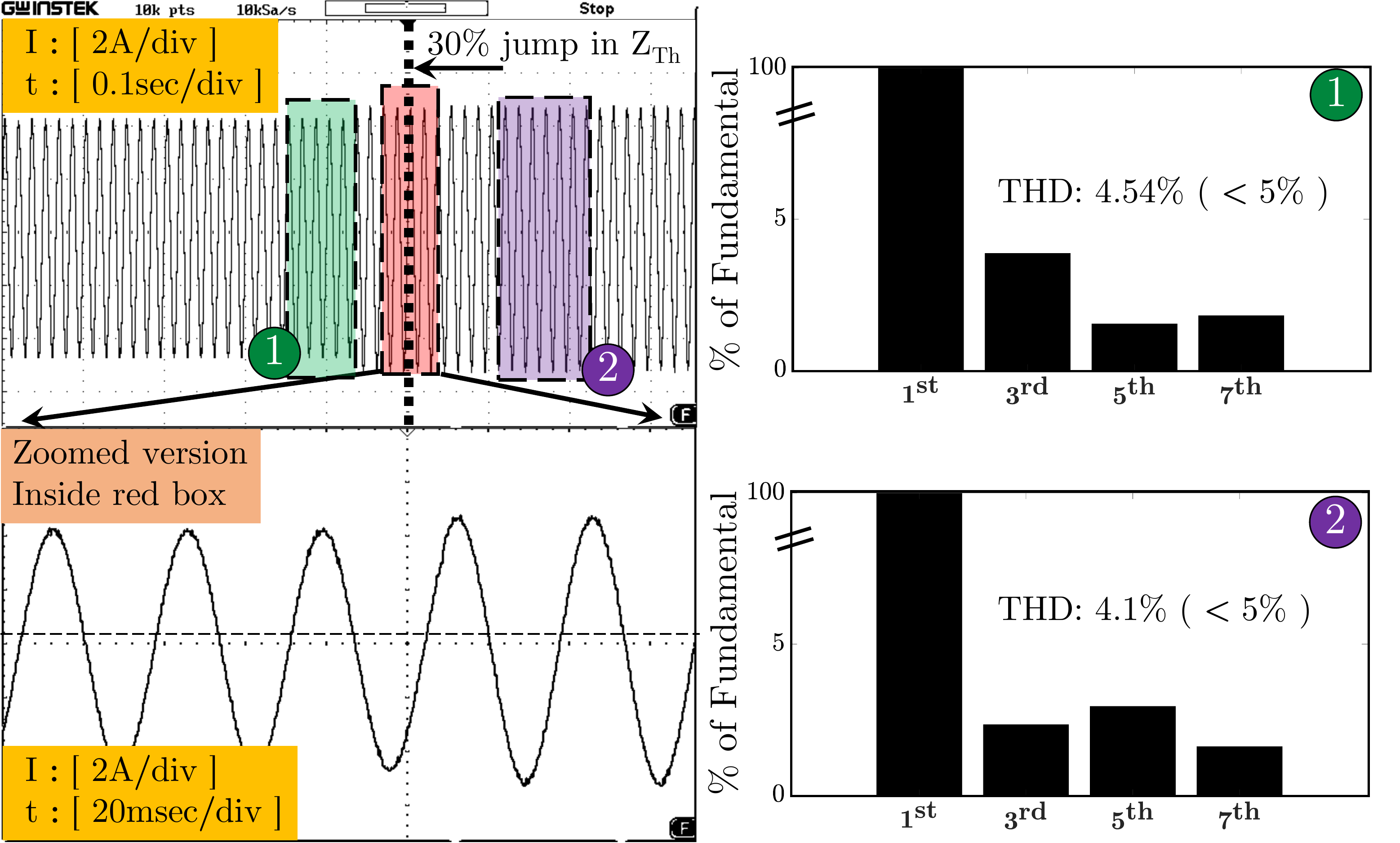}%
	\caption{The current reference tracking capability of GFL inverter in PHIL setup at $\mathrm{Bus}_{10}$ of Fig.~\ref{fig:hardwaresetup}(b) in $\mathtt{CASE}$-$\mathtt{2}$ with the proposed optimal controller.}
	\label{fig:GFLcase2C}
\end{figure}
\subsection{Results and Discussions}
Four test cases (two test cases each for GFL and GFM inverters) are demonstrated by emulating a sequence of events. Two test cases for GFL inverters are as follows:\\
$\bullet~\mathtt{CASE}$-$\mathtt{1}$: The emulated distribution network of Fig.~\ref{fig:hardwaresetup}\subref{fig:PHILsetup} is running in off-grid mode and $\mathrm{P}$-$\mathrm{Q}$ reference of GFL inverters jumps up by $50\%$ due to increased demand.\\
$\bullet~\mathtt{CASE}$-$\mathtt{2}$: The network is running in off-grid mode and experiences a topology change which results in $30\%$ increase in equivalent Thevenin impedance at PCC of GFM inverters.\\
$\bullet~\mathtt{CASE}$-$\mathtt{3}$: The emulated distribution network of Fig.~\ref{fig:hardwaresetup}\subref{fig:PHILsetup} has an on-grid to off-grid mode transition. GFM inverters will have a maximum jump in loading from no-load condition (during on-grid mode) to full-load condition (during off-grid mode).\\
$\bullet~\mathtt{CASE}$-$\mathtt{4}$: The same network has an off-grid to on-grid mode transition. GFM inverters will have another maximum jump in loading from full-load condition (during off-grid mode) to no-load condition (during on-grid mode). Clearly, $\mathtt{CASE}$-$\mathtt{1}$, $\mathtt{CASE}$-$\mathtt{2}$ and $\mathtt{CASE}$-$\mathtt{3}$, $\mathtt{CASE}$-$\mathtt{4}$ are designed in order to capture the robust performance of proposed GFL and GFM control during maximal model uncertainty respectively. 
\begin{figure}[t]
	\centering
	\subfloat[]{\includegraphics[scale=0.11,trim={0cm 0cm 13cm 0cm},clip]{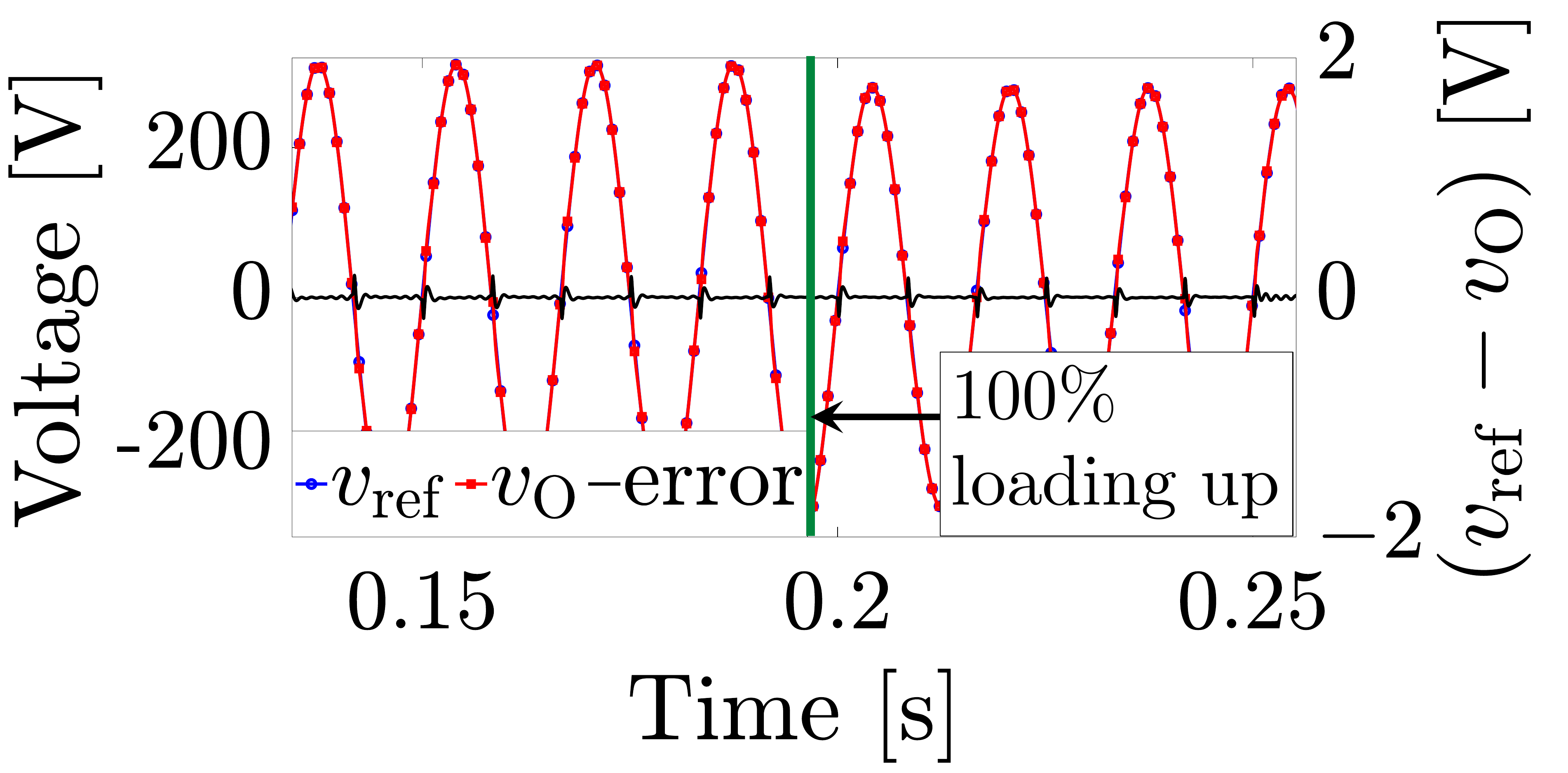}%
	\label{fig:gfm_chil1a}}~
	\subfloat[]{\includegraphics[scale=0.11,trim={0cm 0cm 13cm 0cm},clip]{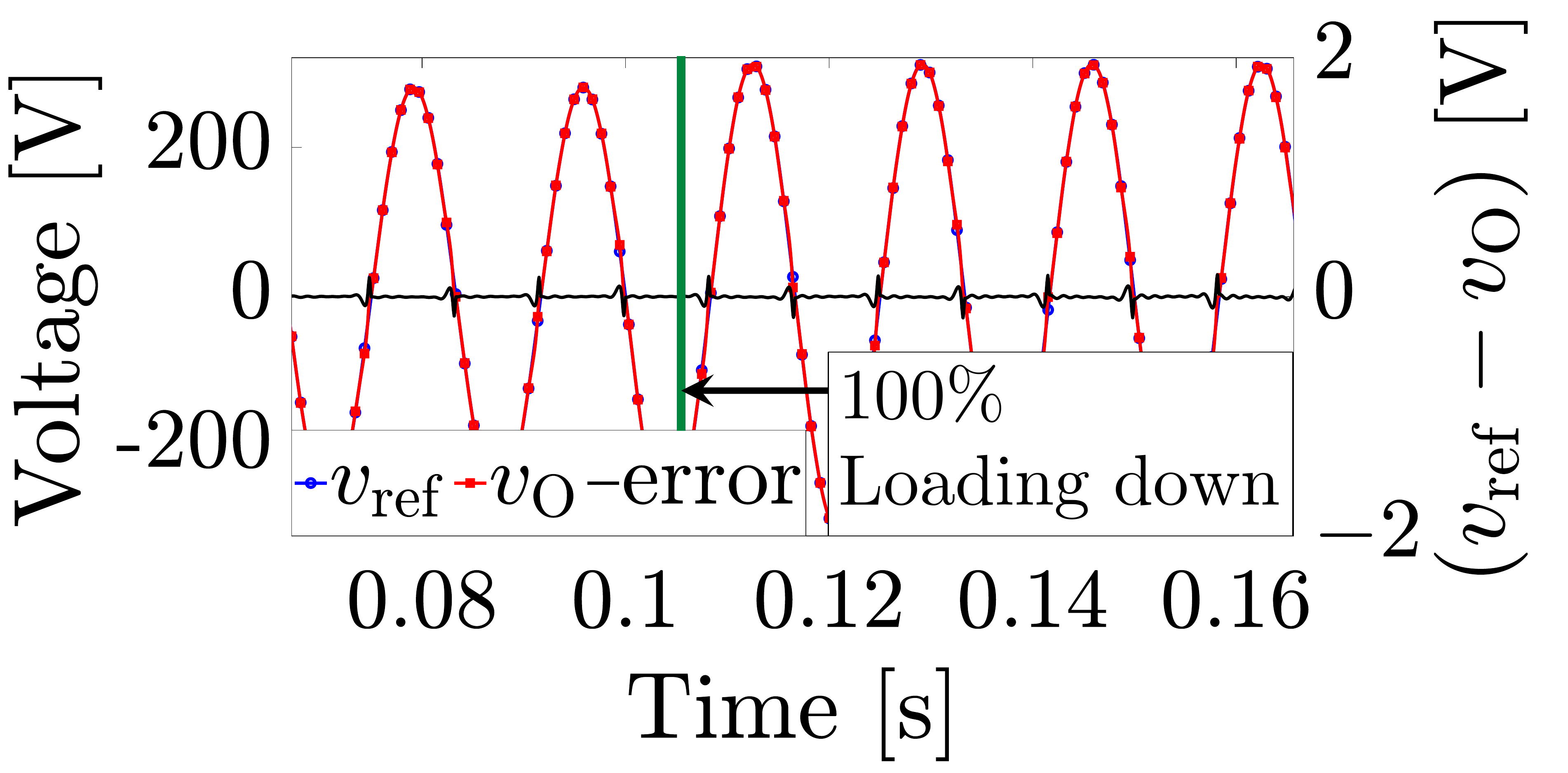}%
	\label{fig:gfm_chil2a}}
	\caption{The voltage reference tracking capability of GFM inverter with the proposed $\mu$-synthesis-based optimal controller in CHIL setup at $\mathrm{Bus}_{10}$ of Fig.~\ref{fig:hardwaresetup}(b) in (a) $\mathtt{CASE}$-$\mathtt{3}$ and (b) $\mathtt{CASE}$-$\mathtt{4}$.}
	\label{fig:gfm_chil1}
\end{figure}
\begin{figure}[t]
	\centering
    \includegraphics[scale=0.21,trim={0cm 0cm 9.5cm 0cm},clip]{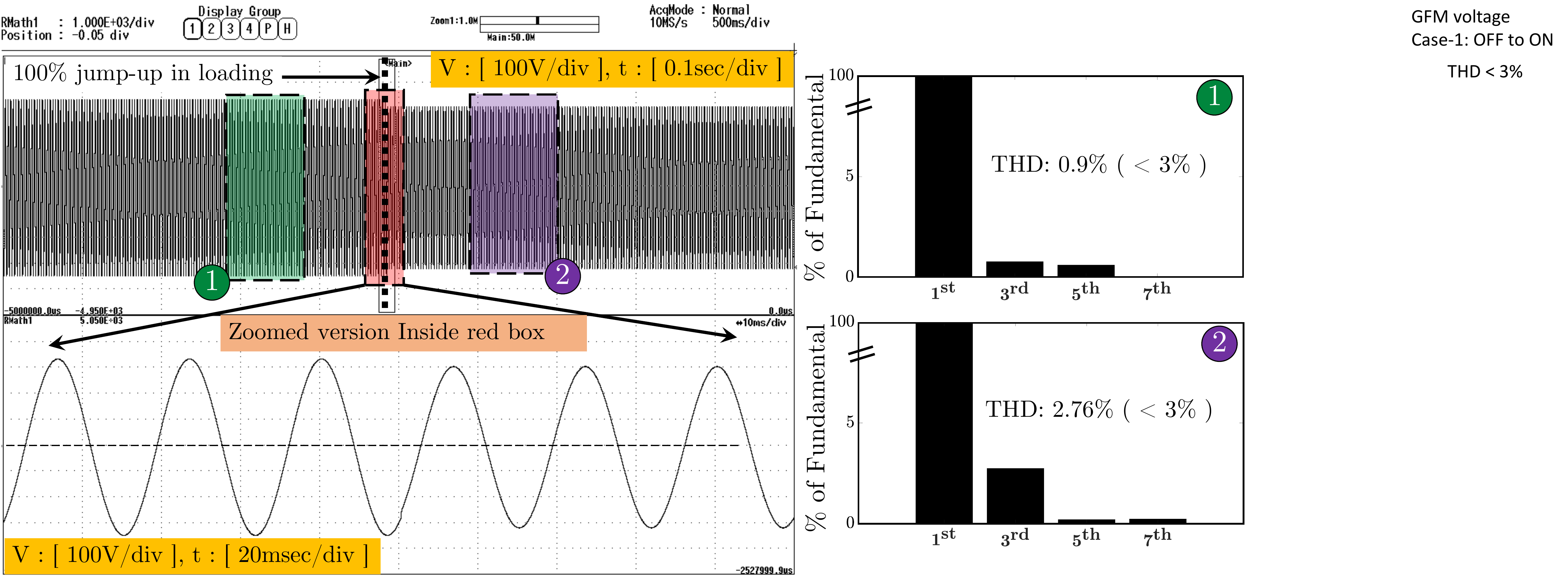}%
	\caption{The voltage reference tracking capability of GFM inverter in PHIL setup at $\mathrm{Bus}_{12}$ of Fig.~\ref{fig:hardwaresetup}(b) in $\mathtt{CASE}$-$\mathtt{3}$ with the proposed optimal controller.}
	\label{fig:GFMcase1V}
\end{figure}
\begin{figure}[t]
	\centering
    \includegraphics[scale=0.21,trim={0cm 0cm 10cm 0cm},clip]{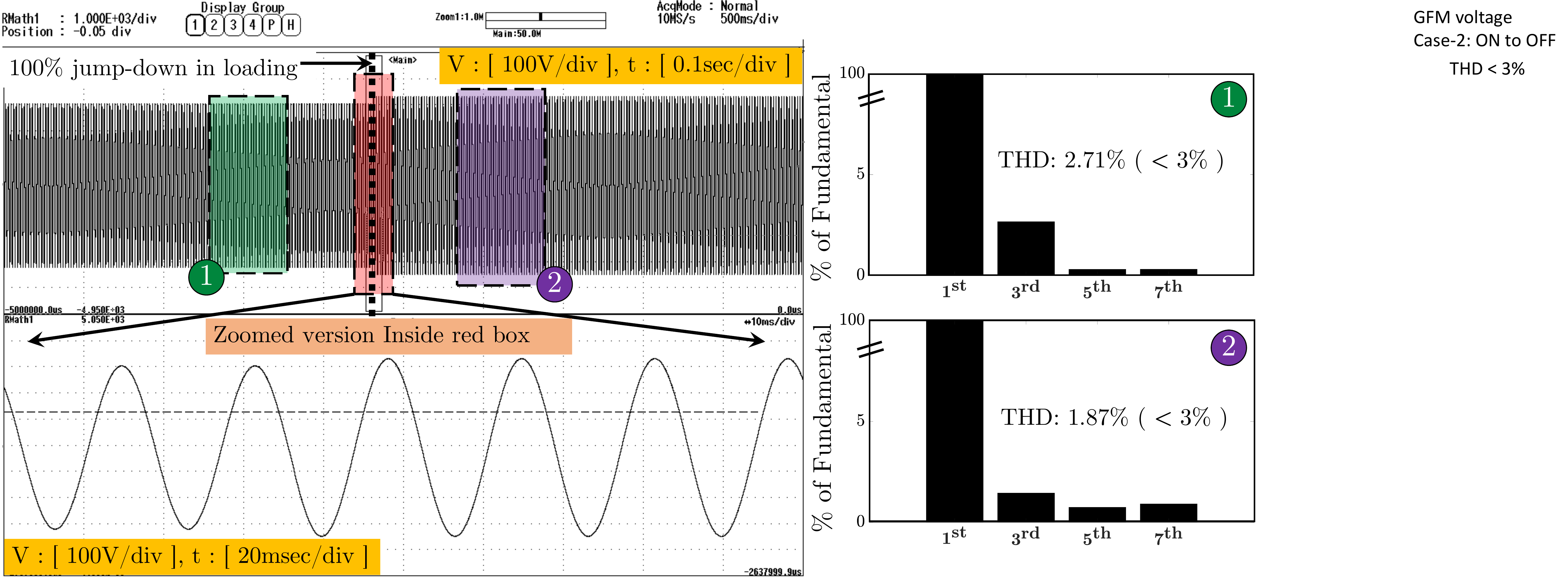}%
	\caption{The voltage reference tracking capability of GFM inverter in PHIL setup at $\mathrm{Bus}_{12}$ of Fig.~\ref{fig:hardwaresetup}(b) in $\mathtt{CASE}$-$\mathtt{4}$ with the proposed optimal controller.}
	\label{fig:GFMcase2V}
\end{figure}
\par Fig.~\ref{fig:gfl_chil1}\subref{fig:gfl_chil1a} and Fig.~\ref{fig:gfl_chil1}\subref{fig:gfl_chil2a} show the current reference tracking capability of the proposed $\mu$-synthesis-based optimal controller for GFL inverter (at $\mathrm{Bus}_{14}$ of Fig.~\ref{fig:hardwaresetup}\subref{fig:PHILsetup}) in CHIL demonstration of $\mathtt{CASE}$-$\mathtt{1}$ and $\mathtt{CASE}$-$\mathtt{2}$ respectively. RMS instantaneous current tracking error ($\mathrm{rICTE}$ in $\%$), defined as $100\times\mathrm{RMS}(i_\mathrm{ref}-i_\mathrm{O})/[\sqrt{2}\mathrm{RMS}(i_\mathrm{ref})]$, is used for assessing the tracking performance of the current controller. It is observed in Fig.~\ref{fig:gfl_chil1}\subref{fig:gfl_chil1a} that both the current reference and output current increases during $50\%$ increase in $\mathrm{P}$-$\mathrm{Q}$ set-points due to adopted reference generation of Appendix~$\mathrm{A}$. The proposed optimal controller has significantly small error in current reference tracking before (\color{black}$\mathrm{rICTE}\approx1.7\%$\color{black}) and after (\color{black}$\mathrm{rICTE}\approx1.9\%$\color{black}) the transition in $\mathtt{CASE}$-$\mathtt{1}$. Similarly, Fig.~\ref{fig:gfl_chil1}\subref{fig:gfl_chil2a} shows that the proposed optimal controller has significantly small error in current reference tracking before (\color{black}$\mathrm{rICTE}\approx1.9\%$\color{black}) and after (\color{black}$\mathrm{rICTE}\approx2.1\%$\color{black}) the jump of equivalent Thevenin impedance in $\mathtt{CASE}$-$\mathtt{2}$. Fig.~\ref{fig:GFLcase1C} shows the current response of GFL inverter (HUT-$2$ at $\mathrm{Bus}_{10}$ of Fig.~\ref{fig:hardwaresetup}\subref{fig:PHILsetup}) as a part of PHIL demonstration of the same event. Here the result is focused on determining the harmonic compensation capability of the proposed optimal controller for GFL inverter during varying reference set-point. It is observed that the total demand distortion (TDD) of current waveform is $<5\%$ before and after the transition as recommended in \cite{ieee2}. Thus, the proposed $\mu$-synthesis-based controller for GFL inverter shows good reference tracking and harmonic compensation capability in $\mathtt{CASE}$-$\mathtt{1}$. Similarly, Fig.~\ref{fig:GFLcase2C} shows the current response of GFL inverter (HUT-$2$ at $\mathrm{Bus}_{10}$ of Fig.~\ref{fig:hardwaresetup}\subref{fig:PHILsetup}) as a part of PHIL demonstration of the same event. Here the result is focused on determining the harmonic compensation capability of the proposed $\mu$-synthesis-based optimal controller for GFL inverter during model uncertainty. It is observed here that the TDD of current waveform is $<5\%$ before and after the transition as recommended in \cite{ieee2}. Thus, the data corroborates the efficacy of the proposed $\mu$-synthesis-based controller for GFL inverter for good reference tracking and harmonic compensation capability in $\mathtt{CASE}$-$\mathtt{2}$. The CHIL and PHIL results substantiate the fact that the proposed $\mu$-synthesis-based optimal controller for GFL is showing robust performance by making sure to have good reference tracking, disturbance rejection and harmonic compensation capability under significant in the plant model uncertainty.
\par Fig.~\ref{fig:gfm_chil1}\subref{fig:gfm_chil1a} and Fig.~\ref{fig:gfm_chil1}\subref{fig:gfm_chil2a} show the voltage reference tracking capability of the proposed $\mu$-synthesis-based optimal controller for GFM inverter (at $\mathrm{Bus}_{10}$ of Fig.~\ref{fig:hardwaresetup}\subref{fig:PHILsetup}) in CHIL demonstration of $\mathtt{CASE}$-$\mathtt{3}$ and $\mathtt{CASE}$-$\mathtt{4}$ respectively. RMS instantaneous voltage tracking error ($\mathrm{rIVTE}$ in $\%$), defined as $100\times\mathrm{RMS}(v_\mathrm{ref}-v_\mathrm{O})/[\sqrt{2}\mathrm{RMS}(v_\mathrm{ref})]$, is used for assessing the tracking performance of the voltage controller. It is observed in Fig.~\ref{fig:gfm_chil1}\subref{fig:gfm_chil1a} that both the voltage reference and output voltage drop during jump in loading (no-load to full-load) due to adopted droop-controlled reference generation of Appendix~$\mathrm{B}$. The proposed $\mu$-synthesis-based optimal controller has significantly small error in voltage reference tracking as shown in Fig.~\ref{fig:gfm_chil1}\subref{fig:gfm_chil1a} before (\color{black}$\mathrm{rIVTE}\approx0.2\%$\color{black}) and after (\color{black}$\mathrm{rIVTE}\approx0.1\%$\color{black}) the increase of equivalent loading. Similarly, it is observed in Fig.~\ref{fig:gfm_chil1}\subref{fig:gfm_chil2a} that both the voltage reference and output voltage, increase during drop in loading (full-load to no-load). The proposed optimal controller has significantly small error in voltage reference tracking before (\color{black}$\mathrm{rIVTE}\approx0.2\%$\color{black}) and after (\color{black}$\mathrm{rIVTE}\approx0.1\%$\color{black}) the decrease in equivalent loading. Fig.~\ref{fig:GFMcase1V} shows the voltage response of GFM inverter (HUT-$1$ at $\mathrm{Bus}_{12}$ of Fig.~\ref{fig:hardwaresetup}\subref{fig:PHILsetup}) as a part of PHIL demonstration of the same event. Here the result is focused on determining the harmonic compensation capability of the proposed optimal controller for GFM inverter during model uncertainty change due to loading. It is observed that the total harmonic distortion (THD) of voltage waveform is $<3\%$ before and after the transition. This is significantly less than the voltage distortion limit ($<8\%$) as recommended in \cite{ieee2}. Thus the data corroborates the advantage of the proposed $\mu$-synthesis-based controller for GFM inverter shows robust performance during model uncertainty caused in $\mathtt{CASE}$-$\mathtt{3}$. Similarly, Fig.~\ref{fig:GFMcase2V} shows the voltage response of GFM inverter (HUT-$1$ at $\mathrm{Bus}_{12}$ of Fig.~\ref{fig:hardwaresetup}\subref{fig:PHILsetup}) as a part of PHIL demonstration of the same event. Here the result is focused on determining the harmonic compensation capability of the proposed optimal controller for GFM inverter during model uncertainty change due to loading. It is observed that the total harmonic distortion (THD) of voltage waveform is $<3\%$ before and after the transition. This is significantly less than the voltage distortion limit ($<8\%$) as recommended in \cite{ieee2}. Thus GFM inverter shows robust performance during model uncertainty caused in $\mathtt{CASE}$-$\mathtt{4}$. The CHIL and PHIL results substantiate the fact that the proposed optimal controller for GFM is showing robust performance by having good reference tracking, disturbance rejection and harmonic compensation capability under significant in the plant model uncertainty.
\begin{figure}[t]
	\centering
	\subfloat[]{\includegraphics[scale=0.11,trim={0cm 0cm 13cm 0cm},clip]{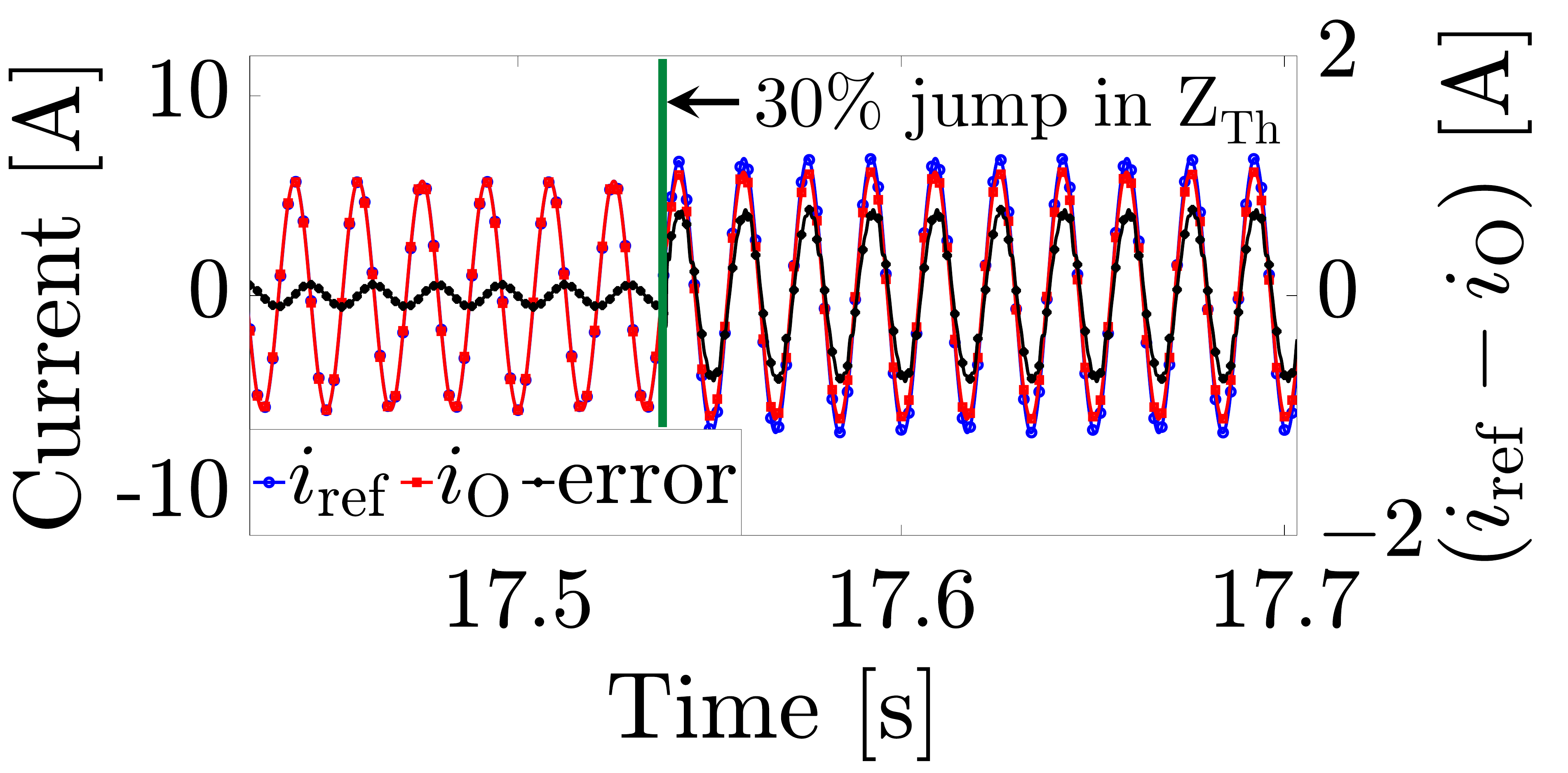}%
	\label{fig:gfl_chil1b}}~
	\subfloat[]{\includegraphics[scale=0.11,trim={0cm 0cm 13cm 0cm},clip]{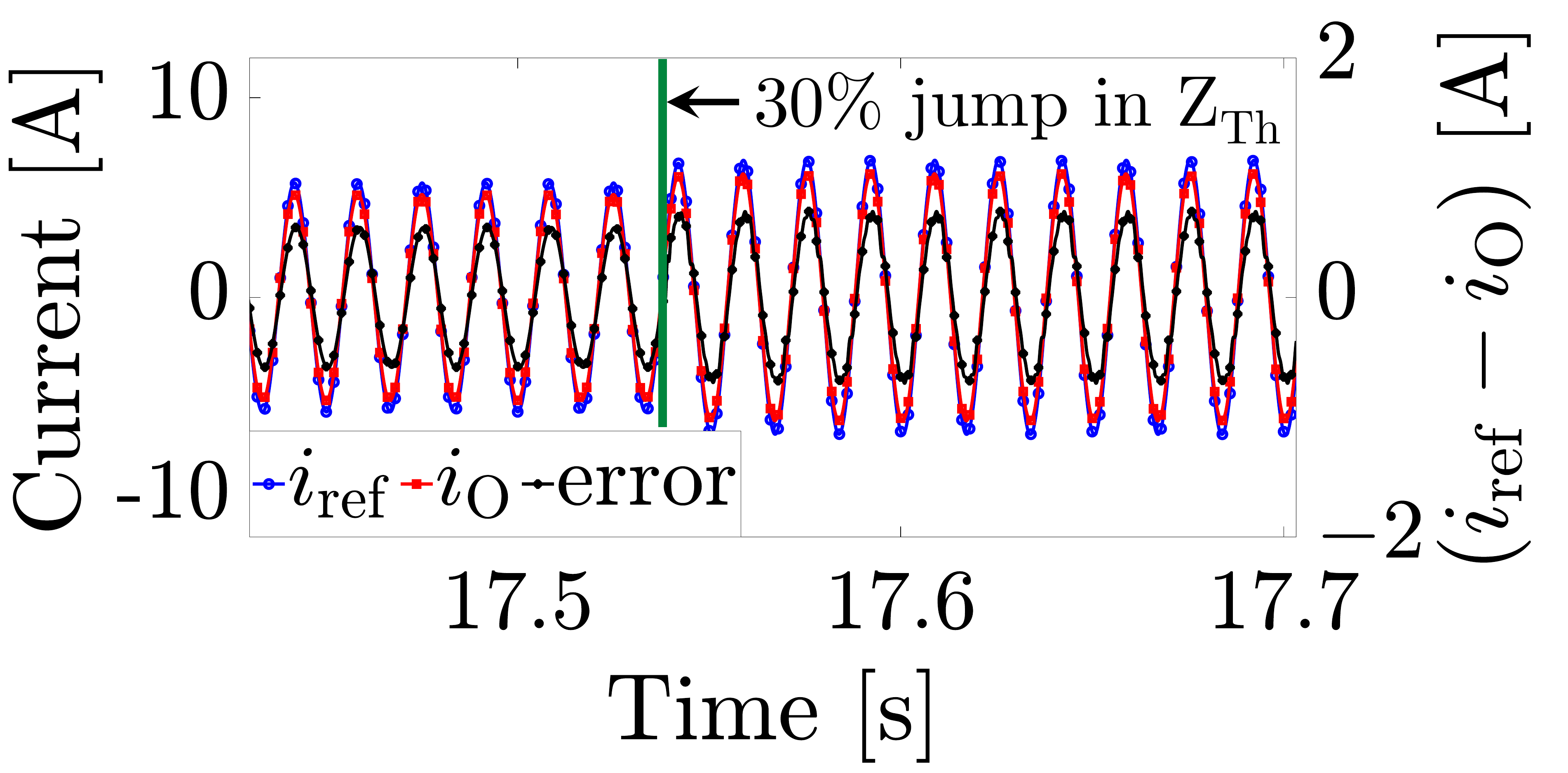}%
	\label{fig:gfl_chil2b}}
	\caption{The current reference tracking capability of GFL inverter in CHIL setup at $\mathrm{Bus}_{14}$ of Fig.~\ref{fig:hardwaresetup}(b) with, (a) $\mathcal{H}_\infty$-based controller without considering uncertainty of Thevenin impedance, (b) the classical PR controller.}
	\label{fig:gfl_chil2}
\end{figure}
\begin{figure}[t]
	\centering
	\subfloat[]{\includegraphics[scale=0.11,trim={0cm 0cm 13cm 0cm},clip]{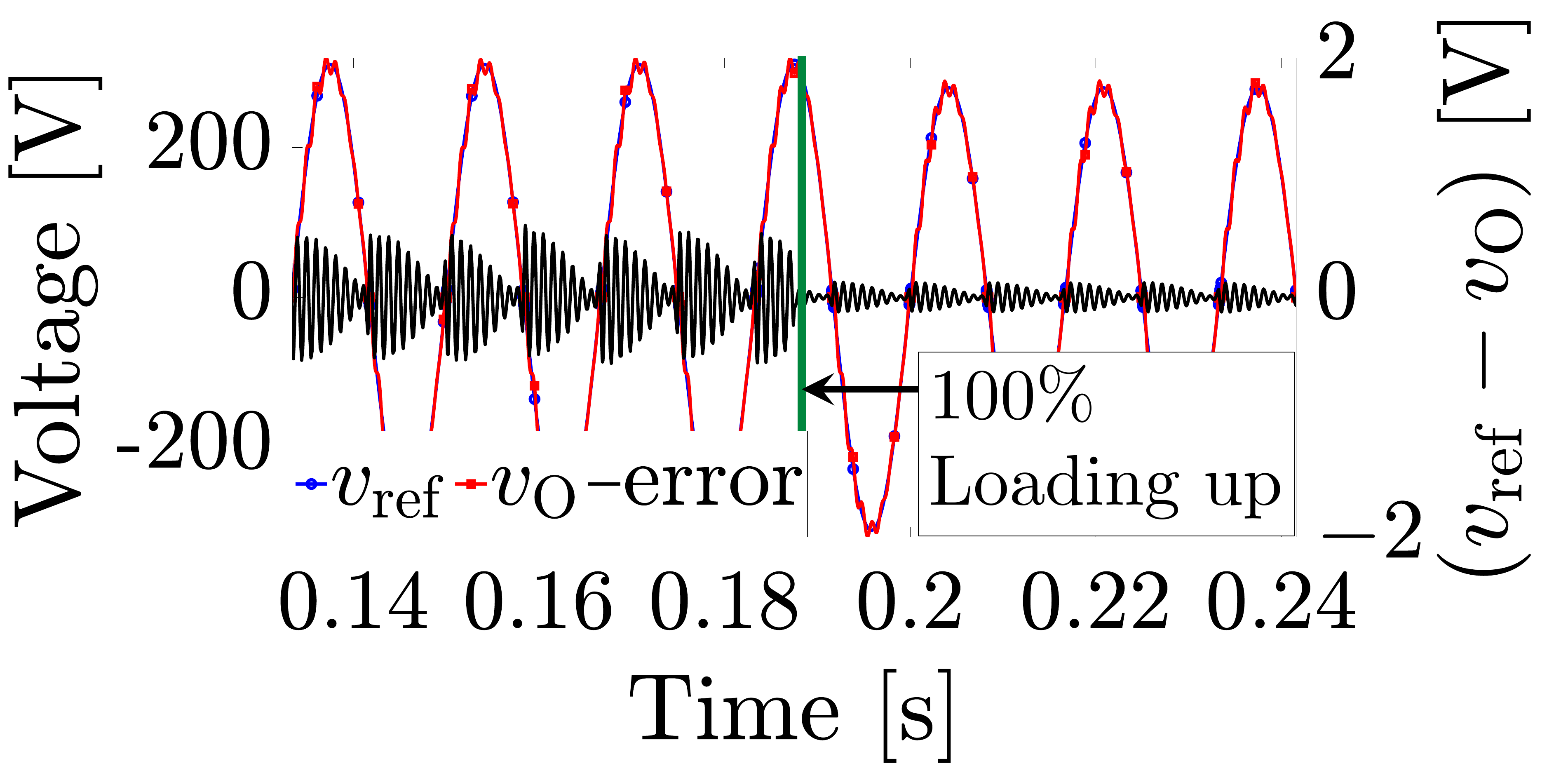}%
	\label{fig:gfm_chil1b}}~
	\subfloat[]{\includegraphics[scale=0.11,trim={0cm 0cm 13cm 0cm},clip]{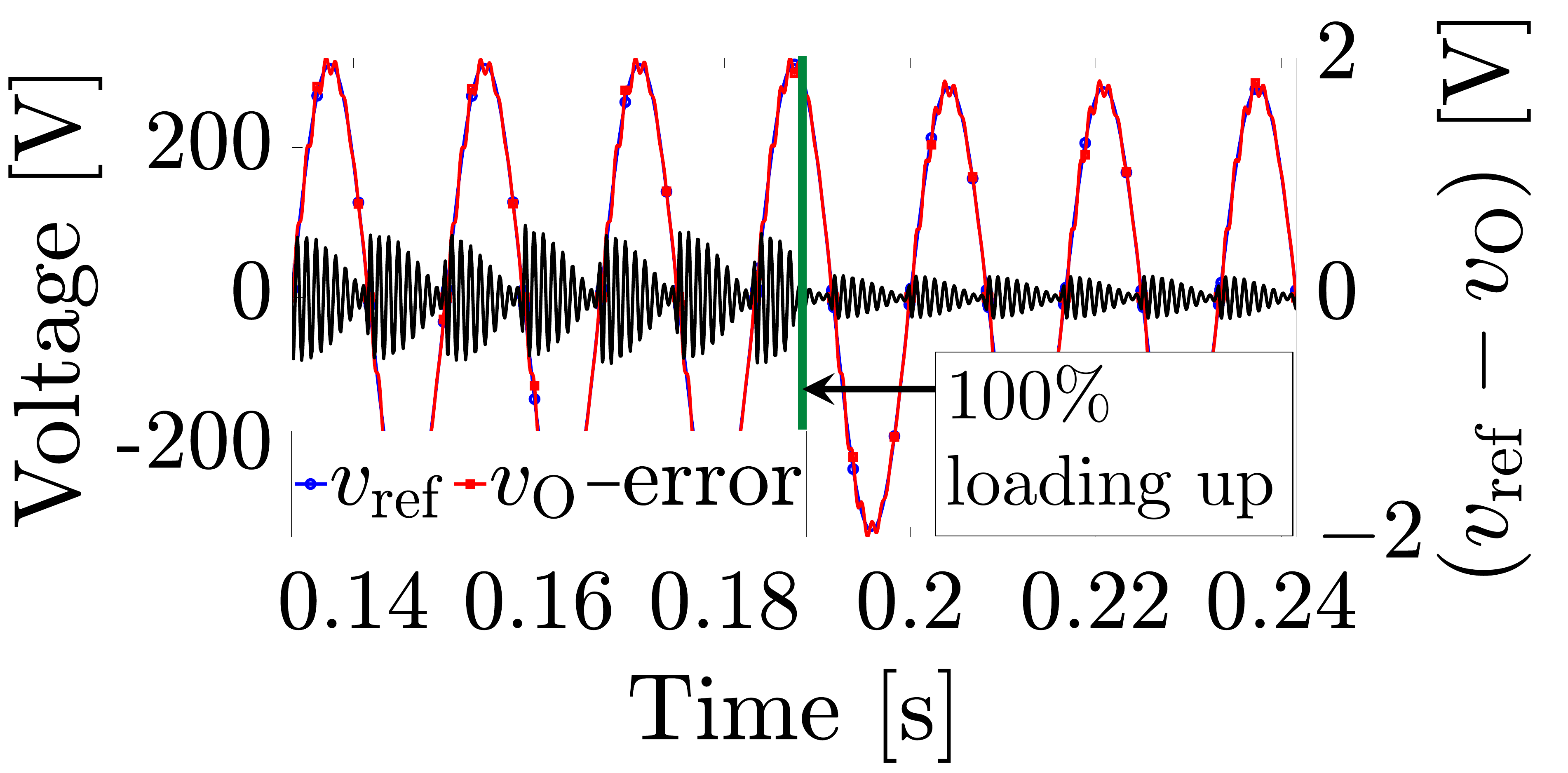}%
	\label{fig:gfm_chil2b}}
	\caption{The voltage reference tracking capability of GFM inverter in CHIL setup at $\mathrm{Bus}_{10}$ of Fig.~\ref{fig:hardwaresetup}(b) with, (a) $\mathcal{H}_\infty$-based controller without considering uncertainty of equivalent loading, (b) the classical PR controller.}
	\label{fig:gfm_chil2}
\end{figure}
\subsection{Performance Comparison}
To showcase the robust performance of the proposed $\mu$-synthesis-based controller, a nominal $\mathcal{H}_\infty$-based robust controller and a classical PR-based controllers are designed for comparison purpose. For GFL inverter, the $\mathcal{H}_\infty$-based robust controller is designed by considering the nominal value of Thevenin impedance. Whereas, for GFM inverter, the $\mathcal{H}_\infty$-based robust controller is designed considering $100\%$ equivalent loading as nominal. Fig.~\ref{fig:gfl_chil2}\subref{fig:gfl_chil1b} and Fig.~\ref{fig:gfl_chil2}\subref{fig:gfl_chil2b} show the current reference tracking capability of the $\mathcal{H}_\infty$-based and PR-based current controller for GFL inverter (at $\mathrm{Bus}_{14}$ of Fig.~\ref{fig:hardwaresetup}\subref{fig:PHILsetup}) in CHIL demonstration of $\mathtt{CASE}$-$\mathtt{2}$ respectively. It is observed that the nominal $\mathcal{H}_\infty$-based robust controller has significant error in current reference tracking ($\mathrm{rICTE}$ increases from $\approx2.5\%$ to $\approx10\%$) after $30\%$ jump in equivalent Thevenin impedance from nominal value. Whereas, the PR-based controller has comparatively larger error both before ($\mathrm{rICTE}\approx6\%$) and after ($\mathrm{rICTE}\approx10\%$) the transition of $\mathtt{CASE}$-$\mathtt{2}$. Whereas, the proposed $\mu$-synthesis-based optimal controller has significantly small error in current reference tracking before (\color{black}$\mathrm{rICTE}\approx1.9\%$\color{black}) and after (\color{black}$\mathrm{rICTE}\approx2.1\%$\color{black}) the jump of equivalent Thevenin impedance in $\mathtt{CASE}$-$\mathtt{2}$ as shown in Fig.~\ref{fig:gfl_chil1}\subref{fig:gfl_chil2a}. Similarly, Fig.~\ref{fig:gfm_chil2}\subref{fig:gfm_chil1b} and Fig.~\ref{fig:gfm_chil2}\subref{fig:gfm_chil2b} show the voltage reference tracking capability of the $\mathcal{H}_\infty$-based and PR-based voltage controller for GFM inverter (at $\mathrm{Bus}_{10}$ of Fig.~\ref{fig:hardwaresetup}\subref{fig:PHILsetup}) in CHIL demonstration of $\mathtt{CASE}$-$\mathtt{3}$ respectively. It is observed that the nominal $\mathcal{H}_\infty$-based robust controller has significant error in voltage reference tracking ($\mathrm{rIVTE}\approx1.2\%$) at no-load condition. Whereas, the PR-based controller has comparatively larger error both before ($\mathrm{rIVTE}\approx1.4\%$) and after ($\mathrm{rIVTE}\approx0.8\%$) the transition of $\mathtt{CASE}$-$\mathtt{3}$. Whereas, the proposed $\mu$-synthesis-based optimal controller has significantly small error in voltage reference tracking before (\color{black}$\mathrm{rIVTE}\approx0.2\%$\color{black}) and after (\color{black}$\mathrm{rIVTE}\approx0.1\%$\color{black}) the increase of equivalent loading as shown in Fig.~\ref{fig:gfm_chil1}\subref{fig:gfm_chil1a}.
%%%%%%%%%%%%%%%%%%%%%%%%%END%%%%%%%%%%%%%%%%%%%%%%%%%%%%%%%%%%
%%%%%%%%%%%%%%%%%%%%%%%%%START%%%%%%%%%%%%%%%%%%%%%%%%%%%%%%%%%%
\section{Conclusion}\label{conclusion}
In this article, a generalized $\mu$-synthesis-based robust control framework is proposed utilizing the fact that there is a voltage-current duality in the plant dynamic model of GFL and GFM inverter. The uncertainties in grid impedance parameters and uncertainties in equivalent loading parameters for GFL and GFM inverters are modeled respectively. The generalized control framework results the controllers that are single-loop, hence simple and cost-effective enough to be implemented, and optimal, in the sense of robustness in performance under uncertainties. The resulting current-controller for GFL inverter provides inherent active damping under grid parameter variation whereas the resulting voltage-controller for GFM inverter enhances the dynamic performance during load transients. A SIL-CHIL-PHIL-based experimental validation evaluates the efficacy and viability of the proposed controllers.
%%%%%%%%%%%%%%%%%%%%%%%%%END%%%%%%%%%%%%%%%%%%%%%%%%%%%%%%%%%%
\section*{APPENDIX}
\subsection*{Appendix~A:~Reference Generation for Grid-following Inverter}
The outer `Reference Generation Block' of Fig.~\ref{fig:GFLGFMckt}\subref{fig:GFLcktv1} eventually generates the $i_\mathrm{ref}$ signal using its pre-specified reference active power, $P_\mathrm{ref}$, and reactive power, $Q_\mathrm{ref}$, (defined locally/centrally) and output signals from phase-locked loop (PLL). The expression of $i_\mathrm{ref}$ is given by
\begin{align}\label{eq:iref}
    i_\mathrm{ref} = \sqrt{2}\frac{\sqrt{P_\mathrm{ref}^2+Q_\mathrm{ref}^2}}{\Tilde{V}}\sin\bigg({\Tilde{\theta}-\arctan{\frac{Q_\mathrm{ref}}{P_\mathrm{ref}}}}\bigg),
\end{align}
where, a $1$-$\phi$ second order generalized integrator-based synchronous reference frame PLL (SOGI-SRF-PLL) operates with its grid-synchronization technique and generates the RMS value, $\Tilde{V}$, and synchronized phase information, $\Tilde{\theta}$, of $v_\mathrm{O}$ \cite{single_phase_pll}.
%%%%%%%%%%%%%%%%%%%%%%%%%%%%%%%%%%%%%%%%%%%%%%%%%%%%%%%%%%%
\subsection*{Appendix~B:~Reference Generation for Grid-forming Inverter}
A $P$-$f$/$Q$-$V$ droop control strategy is adopted for outer `Reference Generation Block' of Fig.~\ref{fig:GFLGFMckt}\subref{fig:GFMcktv1}. The droop characteristic equations are as follows \cite{droop_microgrid_Chandorkar}:
\begin{align}
	\omega &= \omega_\mathrm{N} - n_\mathrm{P}P,~
	V = V_\mathrm{N} - m_\mathrm{Q}Q,\\
	v_\mathrm{ref} &= \sqrt{2}V\sin\bigg(\int\omega\mathrm{d}t\bigg),
\end{align}
where, $\omega_\mathrm{N}$ and $V_\mathrm{N}$ are nominal frequency (in rad/sec) and voltage (RMS) respectively. $P$ and $Q$ are averaged active and reactive power output of GFM inverter. $n_\mathrm{P}$ and $m_\mathrm{Q}$ are the droop coefficients and the values are typically chosen such that $\omega$ and $V$ are within the allowed specification, defined by IEEE $1547$ Standard \cite{ieee1547}, for all $P \in [0, P_\mathrm{rated}]$ and $Q \in [-Q_\mathrm{rated}, Q_\mathrm{rated}]$ respectively. Here, $P_\mathrm{rated}$ and $Q_\mathrm{rated}$ are the rated active and reactive power of the GFM inverter. 
\ifCLASSOPTIONcaptionsoff
  \newpage
\fi
\bibliographystyle{IEEEtran}
\bibliography{biblio}
\end{document}